\def\etal{{\it et~al.}}

\def\gtrapprox{\;\lower 0.5ex\hbox{$\buildrel >
    \over \sim\ $}}             %greater than about
\def\lessapprox{\;\lower 0.5ex\hbox{$\buildrel < \over \sim\ $}}
\documentstyle[11pt,aaspp4]{article}
\begin{document}

\title{A VLA\footnote{The Very Large Array (VLA) is
a facility of the National Radio Astronomy Observatory (NRAO) which is
operated by Associated Universities Inc., under cooperative agreement
with the National Science Foundation.} STUDY OF 15 3CR RADIO GALAXIES}

\author{M\sc ichael H\sc arvanek}
\affil{Center for Astrophysics and Space Astronomy, CB 389
University of Colorado, Boulder, Colorado, 80309-0389, \\
electronic mail: harvanek@casa.colorado.edu}

\and

\author{M\sc artin J. H\sc ardcastle}
\affil{H.H. Wills Physics Laboratory, University of Bristol, Royal Fort,
Tyndall Avenue, Bristol~BS8~1TL, UK \\
electronic mail: M.Hardcastle@Bristol.ac.uk}

% The abstract environment prints out the receipt and acceptance dates
% if they are relevant for the journal style.  For the aasms style, they
% will print out as horizontal rules for the editorial staff to type
% on, so long as the author does not include \received and \accepted
% commands.  This should not be done, since \received and \accepted dates
% are not known to the author.

\begin{abstract}
We present VLA radio maps in total intensity and polarization at 1.4,
4.9 and 8.4 GHz of fifteen 3CR radio galaxies for which good maps
showing the large-scale radio structure have not previously been
available. Previously unknown cores are detected in several sources
and a bright one-sided jet in 3C\,287.1 is mapped for the first time;
several other jet-like features are also imaged. Total and core fluxes
are tabulated and radio core positions are listed and compared to
optical positions.  The galaxy at the optical position listed for
3C\,169.1 is found to lie farther from the radio core position than
another dimmer, bluer galaxy.  We discuss individual sources in some
detail.
\end{abstract}


\section{INTRODUCTION}

This paper is the first in a series dealing with a comparison of
the optical and radio properties of a sample of 3CR radio galaxies
with $0.15 < z < 0.65$.  The sample was taken from the Revised
3C Catalog of Radio Sources of Smith, Spinrad \& Smith (1976) as
updated by \cite{Spin85} and \cite{Spinrad}.  In total, there are
66 objects in the sample and good quality radio maps
showing the large-scale radio structure were found in the literature
for 51 of these objects.  We observed the remaining 15 objects with
the VLA.  In future papers, results of the radio-optical comparison
will be presented.  Here we discuss the new VLA images that were
obtained.

Listed in Table \ref{sample} are the observed sources along with their
optical positions, redshifts, 178 MHz flux densities on the scale of
\cite{Baars}, and spectral indices between 178 and 750 MHz.  (The spectral
index $\alpha$ is defined throughout in the sense $S \propto \nu^{-\alpha}$.)
The optical positions and redshifts are taken from \cite{Spinrad}.
The 178 MHz flux densities are from Kellermann, Pauliny-Toth \& Williams
(1969) except for 3C\,268.2 and 3C\,306.1 which are taken from \cite{Pilk}
and Gower, Scott \& Wills (1967) (both taken from \cite{NED}), respectively,
because their values in \cite{Keller} are likely contaminated with confusing
flux.  All fluxes were adjusted to the scale of \cite{Baars} using the
corrections given in \cite{LaiPea}.  The spectral indices are also consistent
with the scale of \cite{Baars}.  They were taken from Laing, Riley \& Longair
(1983) if available (3C\,16, 3C\,220.1, 3C\,401), or computed using the
178 MHz flux densities given here and 750 MHz flux densities from
\cite{Keller} adjusted to the scale of \cite{Baars} using the corrections
given in \cite{LaiPea}.  Also included in Table \ref{sample} are the rest
frame luminosity at 178 MHz, the scale, the largest angular size (measured
from the best available maps), and the linear size of the source. A
Friedmann cosmology with H$_0$ = 50 km s$^{-1}$ Mpc$^{-1}$ and q$_0$ =
0 was used in the calculation of these quantities and is assumed
throughout this paper.

Table \ref{radio-opt} contains the position of the radio core for
each source, the frequency of the map used to determine the core
position, the reference for the core position, other radio map
references, the optical spectral type of the object (i.e. the ``optical
class'') and the reference
for this type.  The optical type was taken directly from \cite{Jack} who
use slightly different notation than that used here.  Their ``quasar/weak
quasar'' (Q/WQ) classification is equivalent to our broad-line (B) type.
Similarly, their ``high-excitation galaxy'' (HEG) corresponds to our
narrow line (N) type and their ``low-excitation galaxy'' (LEG) is our
low-excitation (E) type.  The references for the optical spectra on which 
these classifications are based can be found in \cite{Jack}.  References
giving additional spectral information that are not found in \cite{Jack}
are included in the Optical Reference column.

\section{THE OBSERVATIONS}

The data presented in this paper are from six separate sets of
observations which are summarized in Table \ref{obs}.  Most observations were
made at L-band in order to map the steep-spectrum extended structure
of the sources.  The C- and X-band observations aid in the detection
and study of the flatter spectrum cores, jets and hot spots.  All
observations were made at two observing frequencies which were later
averaged to give a map at a single effective frequency.  At L-band
these were 1.365 and 1.435 GHz, at C-band these were 4.835 and 4.885
GHz, and at X-band they were 8.415 and 8.465 GHz.  A bandwidth of 50
MHz was used for all observations.

All data were reduced using the AIPS software package in the standard
manner. 3C\,48 and 3C\,286 were used as primary flux and polarization
position angle calibrators.  Bad scans
were removed from the calibrator data until acceptable amplitude and
phase solutions were obtained using nearby point sources as phase
calibrators. These solutions were then applied to the program
sources.
For the L-band data, a small first-order correction for ionospheric
Faraday rotation (corresponding to at most a few degrees at the
frequencies used) was applied using the AIPS task {\sc farad}.
This correction was based on mean monthly sunspot data as
ionospheric measurements were not available.  Finally,
a series of self-calibrations was performed to further improve the
maps. Typically we applied 2 to 3 phase-only self-calibrations
followed by 1 to 2 phase and amplitude self-calibrations.  The
self-calibration procedure was stopped when no significant improvement
in the map was seen.

\section{THE MAPS}

Most maps were made using the AIPS task {\sc imagr}, with robustness
between zero and one (intermediate between uniform and natural
weighting) to improve the noise properties of the maps. For a very few
sources the flux difference between the two observing frequencies
contributed significantly to the off-source noise at 1.4-GHz; in these
cases the two frequencies were mapped separately, a spectral index was
determined, and the visibility amplitudes were corrected to the
average frequency with {\sc imagr} before mapping. The 4.86-GHz map of
3C\,357 was made with AIPS task {\sc apcln}.  The 4.86-GHz map of
3C\,456 was made with AIPS task {\sc mx}.  Both 4.86-GHz maps were
provided courtesy of J.\ Stocke.

Polarization maps as well as total intensity maps are presented for
most sources. However, we have no polarization information for the
objects observed at 1.4-GHz in the A-configuration or for the
4.86-GHz observations because no suitable polarization calibration
data were available. Polarization vectors are only plotted where both
the total and polarized intensities exceed three times the appropriate
r.m.s. noise. The lengths of the vectors on the polarization
maps indicate the degree of polarization; their directions indicate
the directions of the electric field. If the Faraday rotation measure
(RM) along the line of sight is sufficiently low, lines perpendicular
to these vectors would be parallel to the emission-weighted mean
projected direction of the magnetic field vector in the radio
source. The integrated rotation measures towards these sources, where
known, are low (Simard-Normandin, Kronberg \& Button 1981); however,
the rotation measures towards particular regions of the sources are not
known, and so the vector directions must be interpreted with caution.

For two sources, maps of spectral index between 1.4 and 8.44-GHz are
presented. These were made with matching longest and shortest
baselines, and appropriate weighting, to minimize systematic errors
induced by different {\it uv} coverages. Only regions where the signal
exceeded three times the r.m.s. noise on both maps are plotted.

Information concerning the maps presented here is summarized in Table
\ref{maps}.  The dynamic ranges listed in the second to last column are
defined as the peak map flux divided by the r.m.s. noise in the map; the noise
approaches the expected thermal value in many cases in total-intensity
maps and in almost all maps in Stokes Q and U.  The beam size in the last
column is the FWHM along the major and minor axes of the restoring
elliptical Gaussian, obtained by fitting to the dirty beam.  This beam
is shown in one of the corners of each map.  The contour levels and
the location of the peak flux are given in the figure captions.

\section{POLARIZATION}

Since 3C\,348 was observed at two different epochs, a check on the
systematic difference of the 1.4-GHz polarization maps is possible.  A
comparison of the position angles of the polarization maps made at
these two different epochs shows the systematic difference (reflecting
systematics in, among other things, the polarization angle calibration
and the correction for ionospheric Faraday rotation) to be $<$
1$^\circ$.  The r.m.s. is of the order of a few degrees, which is about
what we expect.  Thus the polarization calibration appears to have
worked well.

\section{FLUXES AND POSITIONS}

Total fluxes are given in column 4 of Table \ref{fluxes}. In general, the error
in the total flux is $\simeq$ 5\%.  This includes both the calibration
error and the on-source error.  Column 3 of Table \ref{fluxes} lists the ratio of
the largest angular size of the source to the largest scale structure
visible to the VLA configuration and frequency with which the source
was observed. If this ratio is $\gtrapprox$ 1 then the source may be
undersampled; that is, large scale structure (i.e. structure with
angular size $\gtrapprox$ $\theta_{MAX}$) will be missing from the map
and the total flux presented in Table \ref{fluxes} will be lower than the true value.
Fluxes from individual structures smaller than $\theta_{MAX}$, such as
point sources, are not affected by this undersampling. A ratio $>$ 1
does not necessarily mean that the source is undersampled, however.
For example, a source could be composed of several components that are
all visible to the array but the combined angular size of all these
components is larger than the largest scale structure visible to the
array.  In this case, no structure or flux would be missing from the
map.  For this work, the severity of the undersampling is determined
by comparing the total fluxes from our maps with single-dish total
fluxes from the literature (listed in column 7) and with fully sampled
maps at the same frequency, if available.  Total fluxes in
Table \ref{fluxes} marked with a dagger indicate probable undersampling
and are discussed in the comments on individual sources.  Either these
fluxes are too low to agree (within the errors) with the values in the
literature or, in the case of the 8.44-GHz fluxes, LAS/$\theta_{MAX}$
$>$ 1 and no reliable comparison fluxes are available.
Single-dish fluxes from the literature are presented
in column 7.  If published values agreed, an average value with an
appropriate error is listed.  (If no error was given in the literature,
a value of 5\% was assumed.)  If published values were not in agreement,
a range of values is listed.  This disagreement may be an indication of
source variability.  \cite{Wright} provided an
8.4-GHz comparison flux for 3C\,63 although no error was given.
Other 8.4-GHz comparison fluxes were estimated using
$\alpha^{5000}_{750}$ and $\sim$5-GHz fluxes.  The $\alpha^{5000}_{750}$
values used were those of \cite{Keller} adjusted to the flux scale of
\cite{Baars} using the corrections given in \cite{LaiPea}.

Core flux densities or upper limits are given in column 5 of Table
\ref{fluxes}.  These flux values were determined by one of four
methods.  If the core was visible and relatively isolated, a point
source with characteristics matching those of the beam given in Table
\ref{maps} was fit to the core using the AIPS task {\sc imfit}.  The
integrated flux from this point-source fit is taken to be the core
flux.  If the core was visible but surrounded by confusing flux, a
two-component fit using a point source and an additional sloping
background was performed and the integrated flux from the
point-source fit is quoted in
Table \ref{fluxes}. If the core was not visible, the lobes of the
source were removed by subtracting Gaussian fits, again using {\sc
imfit}.  If the core then became apparent at the appropriate position
(a previously identified radio core position or the optical position
if no core position was available), {\sc imfit} was used to fit a
Gaussian point source to it as above.  Core fluxes obtained in this
manner are marked with an asterisk and an estimate of the uncertainty
of these core fluxes is given below.  If the core still was not
visible, a value of 3 $\times$ the on-source r.m.s. is listed as an
upper limit for the core flux.

The error for the core flux given in column 6 of Table \ref{fluxes} is the value
returned by {\sc imfit}.  This is the flux error due to the
uncertainty of the fit only and does not include the $\sim 5$\% calibration
and on-source error.  For most cases, the error returned by {\sc
imfit} is much larger than 5\% and so this additional error may be
neglected.  However, when the error from {\sc imfit} is $<$ 10\% of
the core flux (3C\,63, 3C\,277, 3C\,287.1, 3C\,357 and 3C\,456), the
additional 5\% error should be included to obtain a more realistic
value.

To estimate the uncertainty of core fluxes measured from
lobe-subtracted maps, the procedure was performed on a 4.86-GHz
B-configuration map of 3C\,456 from \cite{Stocke}.  (This map is not
presented in this paper because the structure is nearly identical to
that in the 1.4-GHz map presented here.)  When compared to the core
flux from the A-configuration map of the same frequency given in Table
\ref{fluxes}, the core flux obtained from the lobe-subtracted map was low by
30\%.  This gives a rough estimate of the uncertainty of the fluxes
determined by this method and indicates that the error returned by
{\sc imfit} may be a substantial underestimate of the true error in
these cases.  However, this source is more complicated and required
the removal of 3 components (see discussion in section \ref{456})
rather than 2 components like the other sources (3C\,320, 3C\,434 and
3C\,459) and so this uncertainty may be an overestimate for these
other sources.  The position of the core obtained from the
lobe-subtracted map was only 0.22$''$ different from that obtained
from the point source fit to the A-configuration map.  This indicates
that the structure remaining after the lobes were subtracted was
indeed the core.

Radio core positions, taken either from our maps or from the literature,
are given in Table \ref{radio-opt}.  Those marked with an asterisk
denote probable cores detected only after the removal of the radio
lobes.  No radio core position is given for
3C\,16 since no core detection is known.  Core positions taken from
maps presented in this paper typically have errors of 0.35$''$ in each
direction.  An examination of the optical positions in Table \ref{sample}
shows that they are within 1$''$ of the radio core positions for most
sources.  Since errors in the optical position are assumed to be a
minimum of 1$''$ in each direction, positions within 1$''$ of each
other are in agreement.  Differences between radio core and optical
positions that are $>$ 1$''$ are discussed in the comments on
individual sources.  Comparisons of different radio core positions for
the same source (i.e. core positions at different frequencies, core
positions from different fits, core positions from the literature) are
also discussed.

\section{COMMENTS ON INDIVIDUAL SOURCES}

\subsection{3C\,16}
(Fig. 1). This source may be undersampled, though the total 8-GHz flux
density is consistent with values predicted from the spectral index and
5-GHz fluxes.  No point-like
core is detected in our 8.44-GHz map and no core detections were found
in the literature.  Two extended knots of emission bracket the
position of the optical identification, which lies on the NE edge of
the southern knot.  Due to the proximity of the optical position to
this knot, the upper limit for the core flux given in Table \ref{fluxes} is a
point source fit to this southern knot, rather than 3 $\times$ the
on-source r.m.s. like the upper limits for the other non-detections.
Neither of these knots is a likely core candidate due to their
extended nature and their steep 1.4--8.4 GHz spectrum ($\alpha \sim
1$).  The two knots may be pieces of an inner twin jet structure, but
their steep spectrum and the fact that they were transversely resolved
by \cite{LP} make this unlikely as well.  An alternative possibility
is that they constitute lobes of a restarting radio source.

\subsection{3C\,63}
(Figs. 2-4). This source is probably undersampled at 1.4-GHz, and the
total flux density is lower than some found in the literature. The
8.44-GHz map is well sampled and the total flux agrees (within the
errors) with the value given in \cite{Wright}.
The similar core fluxes at 1.4 and 8.44-GHz indicate a flat or
slightly inverted spectrum that is typical of a radio core.
The core positions from our two maps agree to within 0.5$''$.

This source was previously observed by \cite{Baum} but the faint
emission regions transverse to the radio axis, particularly
the western region, are more prominent in both our maps.  These
regions constitute a new example of `wings' (cf.\ Leahy \& Perley
1991).  They are strongly polarized (mean fractional polarization of
$\sim$ 50\% at X-band) and the apparent magnetic field is directed
along the wings, transverse to the radio axis, as is typical in such
features.  The spectral index map also shows some interesting
structure.  The hotspots and core have the flattest spectra as
expected, but the steep spectrum of the material around the hot spots
at the sides of the source, particularly in the southern lobe, is
unusual, since aged material is usually found only in the parts of the
lobe nearest the core. This, together with the wings, suggests that
the backflow velocities are very significant compared to hotspot
advance speeds in this source.

\subsection{3C\,169.1}
(Figs. 5-7). This source is probably undersampled at 8.44-GHz
and the total flux density is somewhat lower than
the values predicted from the spectral index and 5-GHz fluxes.
The 1.4-GHz observation is well sampled and our total flux
agrees with published values.

An unresolved radio core is detected in the 8.44-GHz map.  This core
position lies 2.5$''$ NW of the optical position given in Table \ref{sample}.
The errors in this core position are 0.5$''$ in R.A. and 0.2$''$ in
Dec. and the errors in the optical position are given in Kristian,
Sandage \& Katem (1978) as $\sim$ 1$''$ in each direction.  The
combined errors are not large enough to account for the difference in
the two positions.  This is probably due to an underestimate of the
optical position errors.  However, our optical images show another
galaxy closer to the radio core position.  Preliminary measurements
from these images show it to be within 1$''$.  The position for this
second galaxy given in \cite{Krist} (although they refer to it as
``starlike'') puts it 1.3$''$ from the radio core.  This second galaxy
is dimmer (m$_r$ = 21.2 vs. 20.6) and bluer (g-r = 0.5 vs. 0.9) than
the galaxy found at the optical position and, according to \cite{PMC97},
is blueshifted relative to the galaxy found at the optical position
by 1000 km s$^{-1}$.

The 8.44-GHz map shows what appears to be a jet connecting the core to
the northern lobe.  The 1.4-GHz map shows a bulge of emission to the
west, near the center of the source.  The spectral index map shows the
expected steepening back towards the core.  The apparent jet in the
northern lobe has a flatter spectrum than the material around it,
which supports the idea that it actually is a jet rather than an
extension of the lobe.  A higher resolution L-band map which also
shows evidence of the apparent jet can be found in \cite{Neff}.

\subsection{3C\,220.1}
(Fig. 8). The total flux in Table \ref{fluxes} agrees well with values
found in the literature.  The radio core position was measured from a
higher resolution C-band map (Burns \etal\ 1984) and is accurate to
0.5$''$ in each direction.  This core lies 1.9$''$ SE of the optical
position listed in Table \ref{sample}.  The errors in the optical
position are unknown but, assuming they are 1$''$ in each direction,
the combined radio and optical positional errors account for the
difference between the two positions.  Optical images show no objects
closer to the radio core.  The 1.4-GHz map presented here shows no
unusual features in this source. The degree of polarization at this
frequency is unusually low (averaging $\sim 1$\%); this may be
evidence for intrinsic Faraday rotation from the X-ray halo found
around 3C\,220.1 by \cite{HLW98}.

\subsection{3C\,268.2}
(Fig. 9). Although the LAS of this source is 0.94 $\times$
$\theta_{MAX}$ of the configuration used to observe it at 1.4-GHz,
this source may be undersampled as flux densities in the literature
are larger than the total flux density measured from our map. The
radio core coordinates were taken from \cite{Strom} who measured them
from a C-band WSRT map.  This core lies 2.9$''$ NE of the optical
position given in Table \ref{sample}.  No errors are given for the core
coordinates in \cite{Strom} but the errors in the optical position are
listed as 7$''$ in \cite{Wynd} and these are obviously sufficient to
account for the positional difference.  Measurements from optical
images show no objects closer to the radio core and put the core in
the host galaxy (although far from the center).

The lobes in the 1.4-GHz map presented here seem to curve away from
the radio axis near the center of the source.  This behavior is seen
in a number of other sources (e.g. Leahy \& Williams 1984). The heads
of the lobes seem to have a ``pinch'' in the emission located about
one-third of the distance from the leading edge back towards the
center of the source.  This effect is also observed in other radio
galaxies such as 3C\,349 (Hardcastle {\it et al.}\ 1997; Leahy \&
Perley 1991).  A higher resolution L-band map in \cite{Neff} shows
this pinching in the lobe shape more clearly, as well as some details
of the hotspots.

\subsection{3C\,277}
(Fig. 10). Although our map of this giant source may be undersampled,
the total flux density measured from our map is consistent with others
in the literature. The core and a probable jet in the eastern lobe are
visible in our 1.4-GHz map.  Both lobes appear to have double hot
spots.

\subsection{3C\,287.1}
(Fig. 11).  The radio map of this broad-line radio galaxy is almost
certainly undersampled, and the flux density measured from our map is
lower than those found elsewhere. The core flux for this object is
quite large and is a substantial portion (13\%) of the total flux.

The core and a bright, knotty and slightly curved jet in the western
lobe are clearly visible in the 1.4-GHz map. C-band maps can be found
in \cite{Anton} and \cite{Downes} but ours is the first map to show
details of the jet.  The low axial ratio (length to width) of the
source, the prominent core, the one-sided nature of the jet and the
broad-line spectrum (e.g.\ Eracleous \& Halpern 1994) are all
consistent with the object being aligned close to the line of sight.
Hardcastle \etal\ (1998a) suggest that low-power broad-line radio
galaxies of this sort are a low-redshift equivalent of the
high-redshift lobe-dominated quasars, and the radio structure of this
object is certainly consistent with that picture. If the jet's
one-sidedness is caused by relativistic beaming alone, then the
jet-counterjet ratio of $\sim 6$ in the inner 30$''$, corresponding
to the straight part of the jet, implies an angle to the line of sight
of $<70^{\circ}$ and beaming velocities $>0.3c$.

\subsection{3C\,306.1}
(Fig. 12).  This source is not likely to be undersampled.  An
unresolved radio core is detected in the 1.4-GHz map.  This core
position lies 4.2$''$ W and 5.8$''$ S of the optical position listed
in Table \ref{sample}.  Although the errors of the optical position
are given as 7$''$ by \cite{Wynd}, we checked the optical
identification by examining images corresponding to the radio core
position in the Digitized Sky Survey (STScI 1997).  This procedure
returned the same galaxy as that marked on the finding chart of
\cite{Wynd} with coordinates matching those of the radio core to
within 0.4$''$.  Thus, the optical position does indeed coincide with
the radio core.  Our 1.4-GHz map may show some pinching of the
southern lobe (and perhaps the northern lobe to a lesser extent), but
otherwise there is very little distortion in the radio structure.

\subsection{3C\,320}
(Fig. 13). This source is not undersampled. A radio core is not
apparent in our map; however, when the two lobes are removed, a
structure resembling a core remains.  A point source fit to this
structure yields the core flux in Table \ref{fluxes} and the core
position in Table \ref{radio-opt}.  The core is located almost exactly
halfway between the peaks of the two lobes and lies only 1.0$''$ from
the optical position given in Table \ref{sample}.

\subsection{3C\,348}
(Fig. 14).  Also known as Hercules A, this source has one of the
highest measured low-frequency radio flux densities.
We observed this source at two different
epochs.  Although LAS/$\theta_{MAX}$ indicates this source might be
affected by undersampling, the total fluxes in Table \ref{fluxes}
agree well with published values and there does not seem to be much
evidence for missing flux.  This may be a case where the sum of the
angular sizes of all the components making up the source is larger
than $\theta_{MAX}$ but all the individual components are still seen
by the array.  Our two fluxes also agree with each other (within the
errors), as do our two core fluxes; the data from the two epochs are
therefore combined in the map presented here. The core positions from
the two different observations agree to within 0.04$''$.

This source was previously observed at C-band by \cite{Dreher} who
remarked on its peculiar jet-dominated morphology for a source of such
high radio power.  Our 1.4-GHz map shows a weak core as well as the
strong jets.  The extended emission is very symmetric about the
radio axis.  Polarization maps show a strong asymmetry in fractional
polarization, with the western lobe being much less strongly polarized
than the eastern one.  The fact that the more prominent eastern jet
lies in the more strongly polarized lobe combined with the fact that
the source lies in a cluster with strong X-ray emission has led Gizani
\& Leahy (1996, and in prep.) to treat it as an example of the
Laing-Garrington effect (Laing 1988; Garrington {\it et al.}\ 1988).

\subsection{3C\,357}
(Figs. 15-16).  This source is severely undersampled at 4.86-GHz, but
there does not appear to be a problem
at 1.4-GHz.  The radio core lies 1.6$''$ NW of the optical
position listed in Table \ref{sample}.  The optical position is
accurate to 1$''$ in each direction according to
\cite{Veron} and the core position errors are 0.1$''$ in each direction.
When combined, these positional errors account for nearly all the
difference between the two positions.  Measurements from an optical
image put the radio core in the host galaxy (although not at the center).
The positions and fluxes of the cores at 1.4 and 4.86-GHz are
consistent.

The 4.86-GHz map shows multiple hot spots in both lobes. The core is
also visible and a jet points toward the most compact hot spot in the
NW lobe.  The 1.4-GHz map shows some large-scale structure.  The NW
lobe is odd in that the pair of less compact hotspots is
positioned very asymmetrically with respect to the lobe itself and the
axial ratio of this lobe is close to unity.  The SE lobe is extended
to the N, suggesting non-axisymmetric backflow close to the center of
the source.

\subsection{3C\,401}
(Fig. 17).  This source is well sampled by our observations and our
total flux density agrees with the values in the literature. The core
position was measured from a higher resolution 8.44-GHz map in
\cite{Hard} and is accurate to better than 0.1$''$.  The 1.4-GHz core
position is within 1$''$ of this 8.44-GHz core position.  A point
source fit to the radio core seen in the L-band map of \cite{Leahy}
gives an L-band core flux of 21 mJy, consistent with our
measured value.

Our 1.4-GHz map shows both lobes extending away from the radio
axis close to the central regions of the source.  Emission from
the bright one-sided jet is also visible in the southern lobe; this
feature is more clearly seen in the maps of \cite{Hard} and \cite{Leahy}.

\subsection{3C\,434}
(Figs. 18-19).  This source is not undersampled at either observing
frequency and our 1.4-GHz total flux agrees with published values.  A
radio core is not apparent in either map.  However, when the two lobes
are removed from the 8.44-GHz map, a structure resembling a core
remains.  A point source fit to this structure yields the core flux in
Table \ref{fluxes} and the core position in Table \ref{radio-opt}.  The core is located almost
exactly halfway between the peaks of the two lobes and lies only
1.0$''$ from the optical position given in Table \ref{sample}.  When the lobes
are removed from the 1.4-GHz map, no core-like structure is seen.  The
broad lobes seen in the 1.4-GHz map are both slightly extended to the
south near the center of the source.

\subsection{3C\,456}\label{456}
(Figs. 20-21).  The A-configuration 4.86-GHz map presented in
Fig. 21 is definitely undersampled.  A B-configuration 4.86-GHz
map of \cite{Stocke} (LAS/$\theta_{MAX}$ = 0.34) shows large scale
structure almost identical to that in the 1.4-GHz map shown in
Fig. 20 and this structure is clearly missing from the map in Fig. 21.
Due to this undersampling problem, the total 4.86-GHz flux in Table
\ref{fluxes} is taken from the B-configuration map of \cite{Stocke}.
This flux, which is 26\% larger than the total flux measured from
Fig. 21, agrees with values in the literature.  The 1.4-GHz map is
not undersampled and the total flux from this map agrees with
published values.

The radio core position and 4.86-GHz core flux are taken from the map
presented in Fig. 21.  The radio core lies 1.3$''$ SW of the optical
position listed in Table \ref{sample}.  The optical position is
accurate to 1$''$ in each direction according to \cite{Veron} and
these errors account for the difference in the two positions (the
errors in the core position are $<0.1''$).  The core is not apparent
in the 1.4-GHz map but when the lobes are removed (3 Gaussians were
used to remove the lobes - see Fig. 21), a structure resembling a core
remains.  A point source fit to this structure yields the 1.4-GHz core
flux in Table \ref{fluxes} and a core position within 0.2$''$ of that
listed in Table \ref{radio-opt}.  Removing 3 Gaussians from the
B-configuration 4.86-GHz map of \cite{Stocke} and fitting the remaining
structure with a point source yields a core position within 0.3$''$ of
that given in Table \ref{radio-opt} but the core flux in this case is
only 23 mJy, substantially smaller than that measured directly from
Fig. 21.

The 1.4-GHz map shows a small, unusually asymmetrical
source. The faint extension to the east and west of the bright
northern lobe is probably a deconvolution artifact.  The 4.86-GHz
map shows a core with a strong northern lobe or hot spot with a
probable bright backflow feature (the inner northern component)
and a much weaker southern lobe or hot spot.  A possible alternative
interpretation is that the inner northern component is actually
a foreground or background source.  However, optical images show
no evidence of another object near this position.  The extension
to the east of the northern lobe/hotspot in the 4.86-GHz map is
probably a deconvolution artifact as well.

\subsection{3C\,459}
(Fig. 22).  The total flux of this small steep-spectrum source agrees
with values in the literature.  The core is not apparent in this map.
However, when the two lobes are removed, a structure resembling a core
remains.  A point source fit to this structure yields the core flux in
Table \ref{fluxes}. This flux is not consistent with the fluxes
measured by \cite{Ulve} if his suggestion that the core is
steep-spectrum holds at these lower frequencies; however, the position
of the core obtained from this fit lies less than 0.4$''$ from the
core position of \cite{Ulve} given in Table \ref{radio-opt} which was
taken from his high resolution U-band (2 cm) map. The C-band map of
\cite{Morg} appears to be contaminated by imaging artifacts.

\acknowledgments

We thank John Stocke for allowing us to publish his 4.86-GHz maps
and for carefully reading the manuscript and making helpful comments.
M.H. acknowledges the support of a NASA Graduate Student Research
Program Fellowship NGT - 51291.  M.J.H. acknowledges support from
PPARC grant number GR/K98582. This research has made use of the
NASA/IPAC Extragalactic Database (NED) which is operated by the Jet
Propulsion Laboratory, California Institute of Technology, under
contract with NASA.

%\clearpage

% Now comes the reference list.  In this document, we used \cite to call
% out citations, so we must use \bibitem in the reference list, which
% means we use the LaTeX thebibliography environment.  Please note that
% \begin{thebibliography} is followed by a null argument.  If you forget
% this, mayhem ensues, and LaTeX will say "Perhaps a missing item?" when
% you run it.  Do not call us, do not send mail when this happens.  Put
% the silly {} after the \begin{thebibliography}.
%
% Each reference has a \bibitem command to define the citation format
% and the symbolic tag, as well as a \reference command which sets up
% formatting parameters for the reference list itself.
%
% If we had not bothered with the \cite-\bibitem business, calling out
% the references outselves, the reference list could be enclosed in
% a references environment (\begin{references} has no null argument),
% and there would be no need for the leading \bibitem's.
% AAAAAAAAAAAAAA

%\clearpage
%\pagestyle{empty}

\begin{table}
\caption{The Sample of Sources}
\label{sample}
\begin{tabular}{cccccccccc}
\tableline \tableline
Source & R.A. & Dec. & $z$ & $S_{178}$ & $\alpha^{750}_{178}$ & $P_{178}$ (10$^{24}$ & Scale & LAS & Size \\
 & (B1950.) & (B1950.) & & (Jy) &  &  W Hz$^{-1}$ sr$^{-1}$) & (kpc/$''$) & ($''$) & (kpc) \\ \tableline
3C\,16   & 00 35 09.16 &  +13 03 39.6  & 0.406  &  12.2 & 0.94 &  976 & 7.18 &  80.3 &  577 \\
3C\,63   & 02 18 21.90 & $-$02 10 33.  & 0.175  &  20.9 & 0.82 &  252 & 4.01 &  50.0 &  200 \\
3C\,169.1 & 06 47 35.5  &  +45 13 01.  & 0.633  &   8.0 & 0.93 & 1830 & 9.08 &  64.0 &  581 \\
3C\,220.1 & 09 26 31.87 &  +79 19 45.4 & 0.620  &  17.2 & 0.93 & 3750 & 9.00 &  45.4 &  408 \\
3C\,268.2 & 11 58 24.8  &  +31 50 02.  & 0.362  &   7.5 & 0.79 &  441 & 6.70 & 112.7 &  755 \\
3C\,277   & 12 49 26.15 &  +50 50 42.9 & 0.414  &   8.2 & 0.92 &  680 & 7.26 & 159.2 & 1157 \\
3C\,287.1 & 13 30 20.46 &  +02 16 09.0 & 0.2159 &   8.9 & 0.55 &  160 & 4.70 & 139.9 &  658 \\
3C\,306.1 & 14 52 24.5  & $-$04 08 47. & 0.441  &  12.4 & 0.81 & 1150 & 7.53 & 108.0 &  814 \\
3C\,320   & 15 29 29.70 &  +35 43 48.5 & 0.342  &   9.9 & 0.78 &  510 & 6.46 &  37.0 &  239 \\
3C\,348   & 16 48 39.98 &  +05 04 35.0 & 0.154  & 382.6 & 1.03 & 3620 & 3.62 & 202.2 &  732 \\
3C\,357   & 17 26 27.41 &  +31 48 23.9 & 0.1664 &  10.6 & 0.60 &  111 & 3.85 & 118.0 &  454 \\
3C\,401   & 19 39 38.84 &  +60 34 32.6 & 0.201  &  22.8 & 0.71 &  362 & 4.46 &  26.4 &  118 \\
3C\,434   & 21 20 54.40 &  +15 35 11.7 & 0.322  &   5.2 & 0.64 &  226 & 6.22 &  21.6 &  134 \\
3C\,456   & 23 09 56.65 &  +09 03 07.8 & 0.2330 &  11.6 & 0.72 &  253 & 4.97 &  12.2 &   61 \\
3C\,459   & 23 14 02.27 &  +03 48 55.2 & 0.2199 &  27.9 & 0.87 &  554 & 4.77 &  12.6 &   60 \\ \tableline
\end{tabular}
\end{table}

\clearpage

\begin{table}
\caption{Radio Data and Optical Classifications}
\label{radio-opt}
\begin{tabular}{cccccccc}
\tableline \tableline
Source & Core R.A. & Core Dec. & Core Map $\nu$ & Core & Previous Radio & Optical & Optical \\
 & (B1950.) & (B1950.) & (GHz) & Reference & Map Reference & Class & Reference \\ \tableline
 3C\,16   &     ---     &      ---     &   ---   & -- & 6,7,8,9      & N     & 25         \\
 3C\,63   & 02 18 21.94 & $-$02 10 32.7 & 8.440  & 1 & 10,11        & N     & 25         \\
3C\,169.1 & 06 47 35.39 &  +45 13 03.2 &  8.440  & 1 & 12,13        & N     & 25         \\
3C\,220.1 & 09 26 32.29 &  +79 19 43.9 &  4.873  & 2 & 2,9          & N     & 25         \\
3C\,268.2 & 11 58 24.87 &  +31 50 04.8 &  4.874  & 3 & 3,12,14      & N     & 25         \\
3C\,277   & 12 49 26.15 &  +50 50 42.2 &  1.400  & 1 & 3            & E?    & 25         \\
3C\,287.1 & 13 30 20.47 &  +02 16 08.8 &  1.400  & 1 & 15,16,17     & B     & 25,26,27 \\
3C\,306.1 & 14 52 24.22 & $-$04 08 52.8 &  1.400 & 1 & 13           & N     & 25         \\
3C\,320   & 15 29 29.71$^{*}$&  +35 43 49.5$^{*}$&  8.440  & 1 & 18,19  & E & 25         \\
3C\,348   & 16 48 39.98 &  +05 04 35.1 &  1.400  & 1 & 20           & E     & 25       \\
3C\,357   & 17 26 27.30 &  +31 48 24.6 &  4.860  & 1 & 14           & N     & 25         \\
3C\,401   & 19 39 38.82 &  +60 34 32.5 &  8.440  & 4 & 4,21,2,22,9  & E     & 25     \\
3C\,434   & 21 20 54.43$^{*}$&  +15 35 12.6$^{*}$&  8.440  & 1 & 23 & E     & 25         \\
3C\,456   & 23 09 56.59 &  +09 03 06.9 &  4.860  & 1 & 17           & N     & 25         \\
3C\,459   & 23 14 02.31 &  +03 48 55.2 & 14.940  & 5 & 5,11,24      & N     & 25,28    \\ \tableline
\end{tabular}

\tablecomments{Core positions marked with an asterisk denote probable cores detected
after subtraction of the radio lobes.}

\tablerefs{1: This paper;
2: Burns \etal\ 1984;
3: Strom \etal\ 1990;
4: Hardcastle \etal\ 1997;
5: Ulvestad 1985;
6: Leahy \& Perley 1991;
7: Pearson, Perley \& Readhead 1985;
8: Bogers \etal\ 1994;
9: Jenkins, Pooley \& Riley 1977;
10: Baum \etal\ 1988;
11: Rhee \etal\ 1996;
12: Neff, Roberts \& Hutchings 1995;
13: van Breugel 1994;
14: Riley \& Pooley 1975;
15: Antonucci 1985;
16: Downes \etal\ 1986;
17: Stocke 1994;
18: Gregorini \etal\ 1988;
19: Rudnick \& Adams 1979;
20: Dreher \& Feigelson 1984;
21: Leahy, Bridle \& Strom 1997;
22: Laing 1981;
23: Pooley \& Henbest 1974;
24: Morganti, Killeen \& Tadhunter 1993;
25: Jackson \& Rawlings 1997 and references therein;
26: Antonucci 1982;
27: Grandi \& Osterbrock 1978;
28: Tadhunter \etal\ 1993.}

\end{table}

\clearpage

\begin{table}
\caption{Observations}
\label{obs}
\begin{tabular}{ccccccc}
\tableline \tableline
 & & VLA & $\nu$ & $\lambda$ & & Duration\\
Source & Date & Configuration & (GHz) & (cm) & Band & (min) \\ \tableline
 3C\,16   & 1994 Oct 22 & C & 8.4399 &  3.6 & X & 47 \\
 3C\,63   & 1995 Aug 11 & A & 1.4000 & 20   & L & 32 \\
          & 1994 Oct 22 & C & 8.4399 &  3.6 & X & 40 \\
3C\,169.1 & 1995 Oct 09 & B & 1.4000 & 20   & L & 15 \\
          & 1994 Oct 22 & C & 8.4399 &  3.6 & X & 40 \\
3C\,220.1 & 1995 Oct 09 & B & 1.4000 & 20   & L & 18 \\
3C\,268.2 & 1995 Oct 09 & B & 1.4000 & 20   & L & 36 \\
3C\,277   & 1995 Oct 09 & B & 1.4000 & 20   & L & 36 \\
3C\,287.1 & 1996 Jan 08 & B & 1.4000 & 20   & L & 66 \\
3C\,306.1 & 1995 Oct 09 & B & 1.4000 & 20   & L & 36 \\
3C\,320   & 1994 Oct 22 & C & 8.4399 &  3.6 & X & 20 \\
3C\,348   & 1995 Oct 09 & B & 1.4000 & 20   & L & 36 \\
          & 1996 Jan 08 & B & 1.4000 & 20   & L & 36 \\
3C\,357   & 1995 Oct 09 & B & 1.4000 & 20   & L & 18 \\
          & 1984 Feb 02 & B & 4.8600 &  6   & C & 39 \\
3C\,401   & 1995 Aug 11 & A & 1.4000 & 20   & L & 32 \\
3C\,434   & 1995 Aug 11 & A & 1.4000 & 20   & L & 24 \\
          & 1994 Oct 22 & C & 8.4399 &  3.6 & X & 40 \\
3C\,456   & 1995 Aug 11 & A & 1.4000 & 20   & L & 32 \\
          & 1985 Mar 03 & A & 4.8600 &  6   & C & 29 \\
3C\,459   & 1995 Aug 11 & A & 1.4000 & 20   & L & 32 \\ \tableline
\end{tabular}
\end{table}

\clearpage

\begin{table}
\caption{Total Intensity Maps}
\label{maps}
\begin{tabular}{ccccccc}
\tableline \tableline
 & $\nu$ & Peak Flux & Map r.m.s. & Theoretical r.m.s. & Dynamic & Beam\\
Source & (GHz) & (mJy/Beam) & ($\mu$Jy) & ($\mu$Jy) & Range & ($''$) $\times$ ($''$) @ P.A. ($^\circ$)\\ \tableline
 3C\,16   & 8.44 &   28.9 &  26 & 21 & 1100 & 2.47 $\times$ 2.34 @  $-12.98$ \\
 3C\,63   & 1.4  &  250 & 161 & 34 & 1600 & 2.30 $\times$ 1.43 @ $-43.03$ \\
         & 8.44 &   81.3 &  25 & 22 & 3300 & 3.29 $\times$ 2.40 @ $-$37.65 \\
3C\,169.1 & 1.4  &  287 & 144 & 49 & 2000 & 7.10 $\times$ 4.28 @  83.66 \\
         & 8.44 &   41.3 &  28 & 22 & 1500 & 3.85 $\times$ 2.20 @ $-$78.46 \\
3C\,220.1 & 1.4  &  596 & 213 & 44 & 2800 & 6.10 $\times$ 3.92 @ $-$61.24 \\
3C\,268.2 & 1.4  &  182 & 112 & 31 & 1600 & 4.60 $\times$ 4.29 @  $-82.59$ \\
3C\,277   & 1.4  &   93.3 &  74 & 31 & 1300 & 4.75 $\times$ 4.33 @ $-$29.97 \\
3C\,287.1 & 1.4  &  468 & 125 & 23 & 3700 & 4.99 $\times$ 4.45 @   8.62 \\
3C\,306.1 & 1.4  &  548 & 246 & 31 & 2200 & 5.45 $\times$ 4.51 @  $-7.02$ \\
3C\,320   & 8.44 &   94.0 &  40 & 31 & 2400 & 6.72 $\times$ 2.13 @  64.40 \\
3C\,348   & 1.4  & 2090 & 543 & 31 & 3800 & 5.68 $\times$ 4.56 @ $-41.58$ \\
3C\,357   & 1.4  &  177 & 129 & 44 & 1400 & 5.25 $\times$ 4.33 @ $-83.03$ \\
         & 4.86 &   11.7 & 170 & 31 &   70 & 1.79 $\times$ 1.12 @ $-88.96$ \\
3C\,401   & 1.4  &  236 & 104 & 34 & 2300 & 1.65 $\times$ 1.20 @ $-56.87$ \\
3C\,434   & 1.4  &  136 &  66 & 38 & 2100 & 1.47 $\times$ 1.39 @  0.69\\
         & 8.44 &   57.8 &  29 & 22 & 2000 & 2.62 $\times$ 2.47 @  65.19 \\
3C\,456   & 1.4  & 1240 & 181 & 34 & 6900 & 1.49 $\times$ 1.31 @  $-3.52$ \\
         & 4.86 &  237.9 & 423 & 32 &  560 & 0.54 $\times$ 0.44 @ $-55.53$ \\
3C\,459   & 1.4  & 2370 & 305 & 34 & 7800 & 1.57 $\times$ 1.33 @  $-3.92$ \\ \tableline
\end{tabular}
\end{table}

\clearpage

\begin{table}
\caption{Fluxes}
\label{fluxes}
\begin{tabular}{cccccccc}
\tableline \tableline
Source & $\nu$ & LAS/$\theta_{MAX}$ & Total Flux & Core Flux & Error & Single-Dish & Single-Dish \\
 & (GHz) & & (Jy) & (mJy) & (mJy) & Flux (Jy) & Flux Reference \\ \tableline
 3C\,16   & 8.44 & 1.34 &  0.290$\dag$ &  $<$ 3.5 &  & (0.26-0.30) & (1,2,3,4) \\
 3C\,63   & 1.4  & 1.32 &  3.50$\dag$  &  16   &  1 & 3.3-3.9 & 5,3,4 \\
         & 8.44 & 0.83 &  0.438  &  18   &  1 & 0.46 & 3 \\
3C\,169.1 & 1.4  & 0.53 &  1.28   &  $<$ 6   &  & 1.27 $\pm$ 0.07 & 5,4 \\
         & 8.44 & 1.07 &  0.204$\dag$ &      1.0 &  0.1 & (0.21-0.25) & (1,2,4) \\
3C\,220.1 & 1.4  & 0.38 &  2.27   & $<$ 20   &  & 2.27 $\pm$ 0.08 & 5,4 \\
3C\,268.2 & 1.4  & 0.94 &  1.28$\dag$  & $<$ 10   &  & 1.51 $\pm$ 0.09 & 5,4 \\
3C\,277   & 1.4  & 1.33 &  1.16$\dag$  &      3.6 &  0.1 & 1.10-1.34 & 5,4 \\
3C\,287.1 & 1.4  & 1.17 &  2.63$\dag$  &    333   &  2 & 2.8-3.1 & 5,3,6,4 \\
3C\,306.1 & 1.4  & 0.90 &  2.09   &      5.7 &  0.8 & 1.90 $\pm$ 0.08 & 5,3,4 \\
3C\,320   & 8.44 & 0.62 &  0.277  &      6$^{*}$  &  1 & (0.30-0.32) & (1,2,4) \\
3C\,348   & 1.4  & 1.69 & 48.0    &     62   &  8 & 44.9-48.6 & 5,3,6,4 \\
         & 1.4  & 1.69 & 46.6    &     55   &  6 &  &  \\
3C\,357   & 1.4  & 0.98 &  2.77   &   9   &  1 & 2.7 $\pm$ 0.1 & 5,6,4 \\
         & 4.86 & 3.28 &  0.296$\dag$ &      6.5 &  0.3 & 0.97 $\pm$ 0.05 & 1,2,6,4 \\
3C\,401   & 1.4  & 0.69 &  5.01   &  21   &  4  & 4.8-5.4 & 5,6,4 \\
3C\,434   & 1.4  & 0.57 &  1.34   &  $<$ 9   &  & 1.28 $\pm$ 0.05 & 5,3,4 \\
         & 8.44 & 0.36 &  0.284  &      8$^{*}$  &  1  & (0.24-0.35) & (1,2,3,4) \\
3C\,456   & 1.4  & 0.32 &  2.55   &     38$^{*}$  &  3  & 2.56 $\pm$ 0.07 & 5,3,4 \\
         & 4.86 & 1.22 &  0.786\tablenotemark{a} &     30   &  1 & 0.67-0.80 & 7,1,2,3,4 \\
3C\,459   & 1.4  & 0.33 &  4.56   &     99$^{*}$  & 23 & 4.1-4.7 & 5,3,6,4 \\ \tableline
\end{tabular}

\tablenotetext{\rm a}{This flux is taken from a B-configuration map of
Stocke 1994 that is well sampled (LAS/$\theta_{MAX}$ = 0.34).\\  }

\tablecomments{Total fluxes marked with a dagger are possibly affected by
undersampling. Core fluxes marked with an asterisk are determined
after subtraction of the radio lobes, modelled as Gaussian components.
Single-dish fluxes in parentheses are estimated values obtained from
$\sim$5-GHz fluxes from the references in parentheses and
$\alpha^{5000}_{750}$ values adjusted to the flux scale of
Baars \etal\ 1977.}

\tablerefs{1: Gregory \& Condon 1991;
2: Becker, White \& Edwards 1991;
3: Wright \& Otrupcek 1990;
4: Kellermann \etal\ 1969 adjusted to the flux scale of
Baars \etal\ 1977 using the corrections given in Laing \& Peacock 1980;
5: White \& Becker 1992;
6: K\"uhr \etal\ 1981;
7: Griffith \etal\ 1995.}

\end{table}

\clearpage

% Finally, we have figure captions.  Usually these must be on a separate
% page, although unlike table, it is often permissible to have several
% figure captions on the same page.  We force the page break between
% the reference list and the figure captions.
%
% The \caption command in the figure environment works like the one in the
% table environment (it's the same one, actually), except that this one
% produces identification text that reads "Figure N."

\begin{figure}
\plottwo{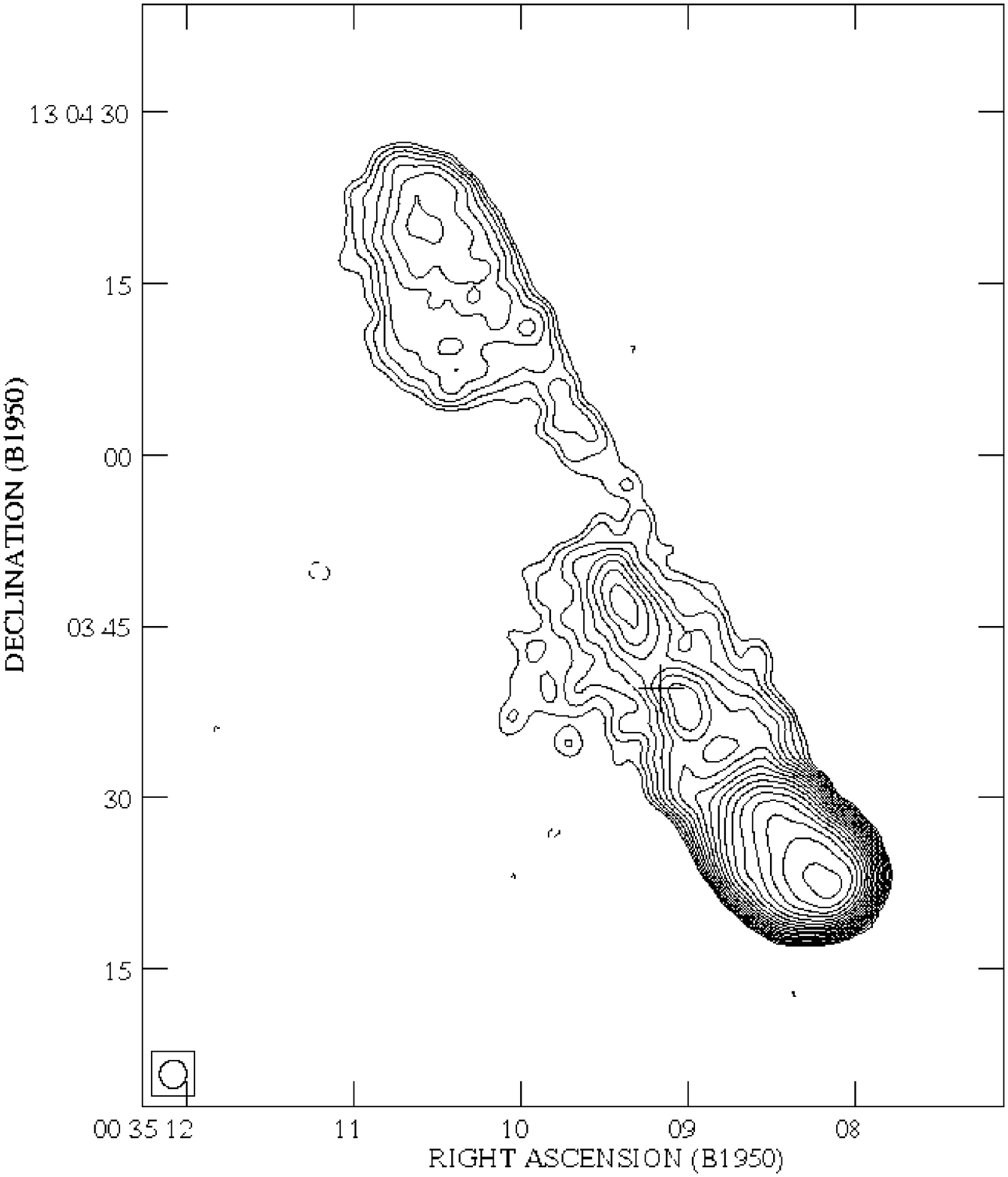}{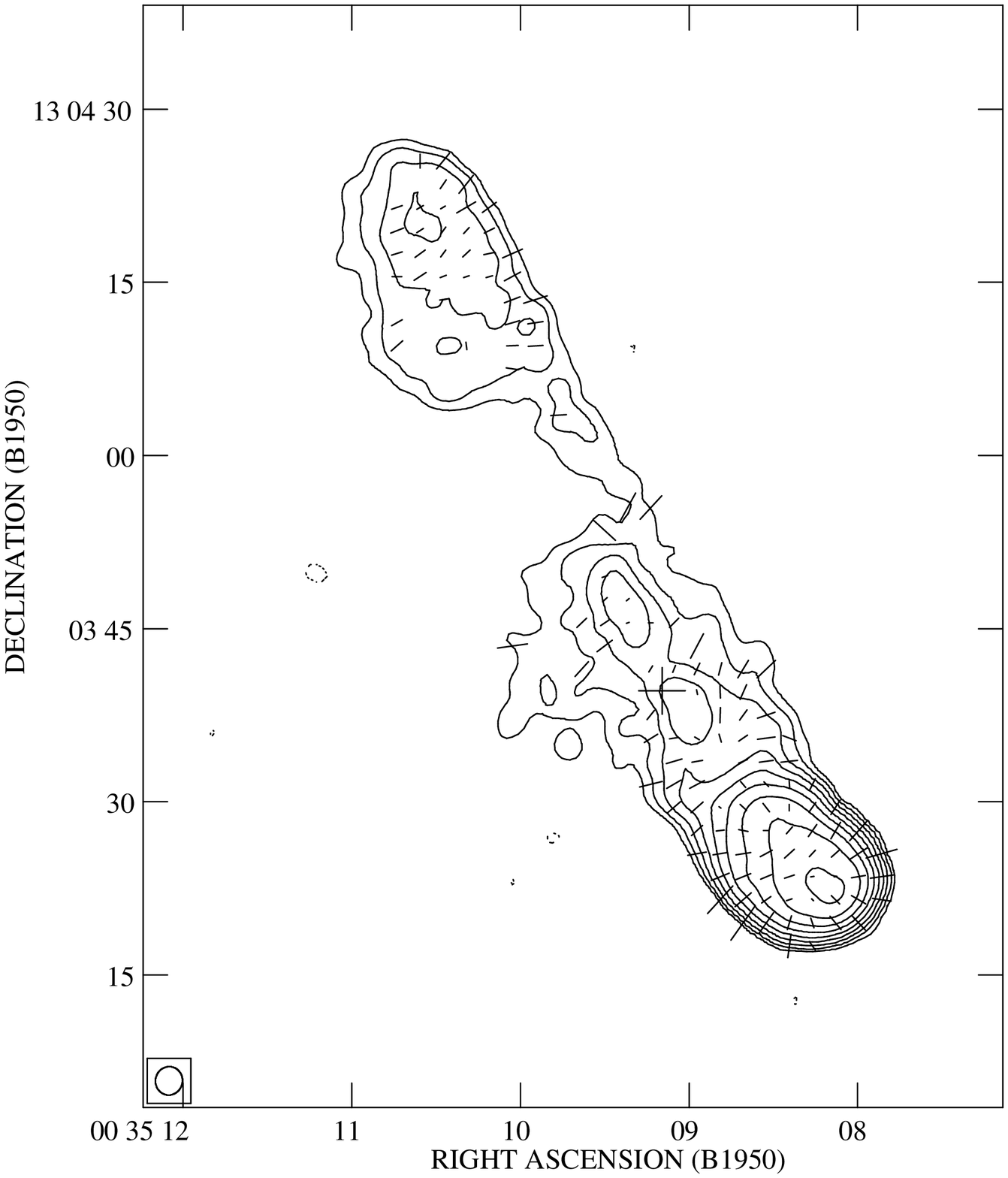}
\figcaption{Left: Total intensity map of 3C\,16 at 8.44 GHz.
The contour levels are $0.1 \times (-\protect\sqrt 2, -1, 1,
\protect\sqrt 2, 2, 2\protect\sqrt 2, \dots)$ mJy beam$^{-1}$. Right:
Polarization map. The contour levels are $0.1 \times (-2, -1, 1, 2, 4,
\dots)$ mJy beam$^{-1}$. A vector of length one arcsecond corresponds
to 20\% polarization. A cross marks the position of the optical
identification. The peak flux occurs in the southern lobe.}
\end{figure}

\begin{figure}
%\epsscale{0.5}
\plotone{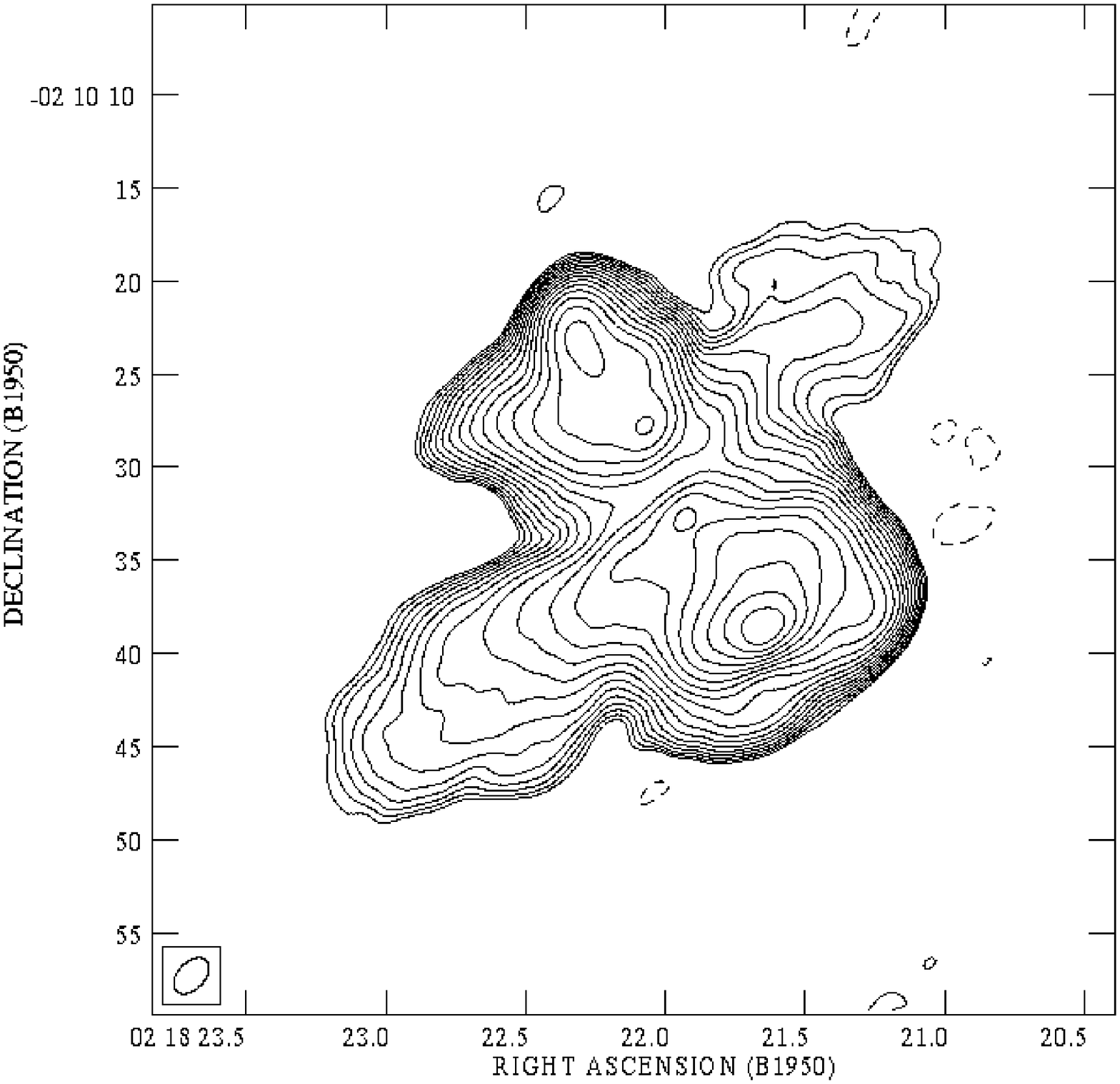}
\figcaption{Total intensity map of 3C\,63 at 1.4 GHz.
The contour levels are $0.5 \times (-\protect\sqrt 2, -1, 1,
\protect\sqrt 2, 2, 2\protect\sqrt 2, \dots)$ mJy beam$^{-1}$. The
peak flux occurs in the southern lobe.}
%\epsscale{1.0}
\end{figure}

\begin{figure}
\plottwo{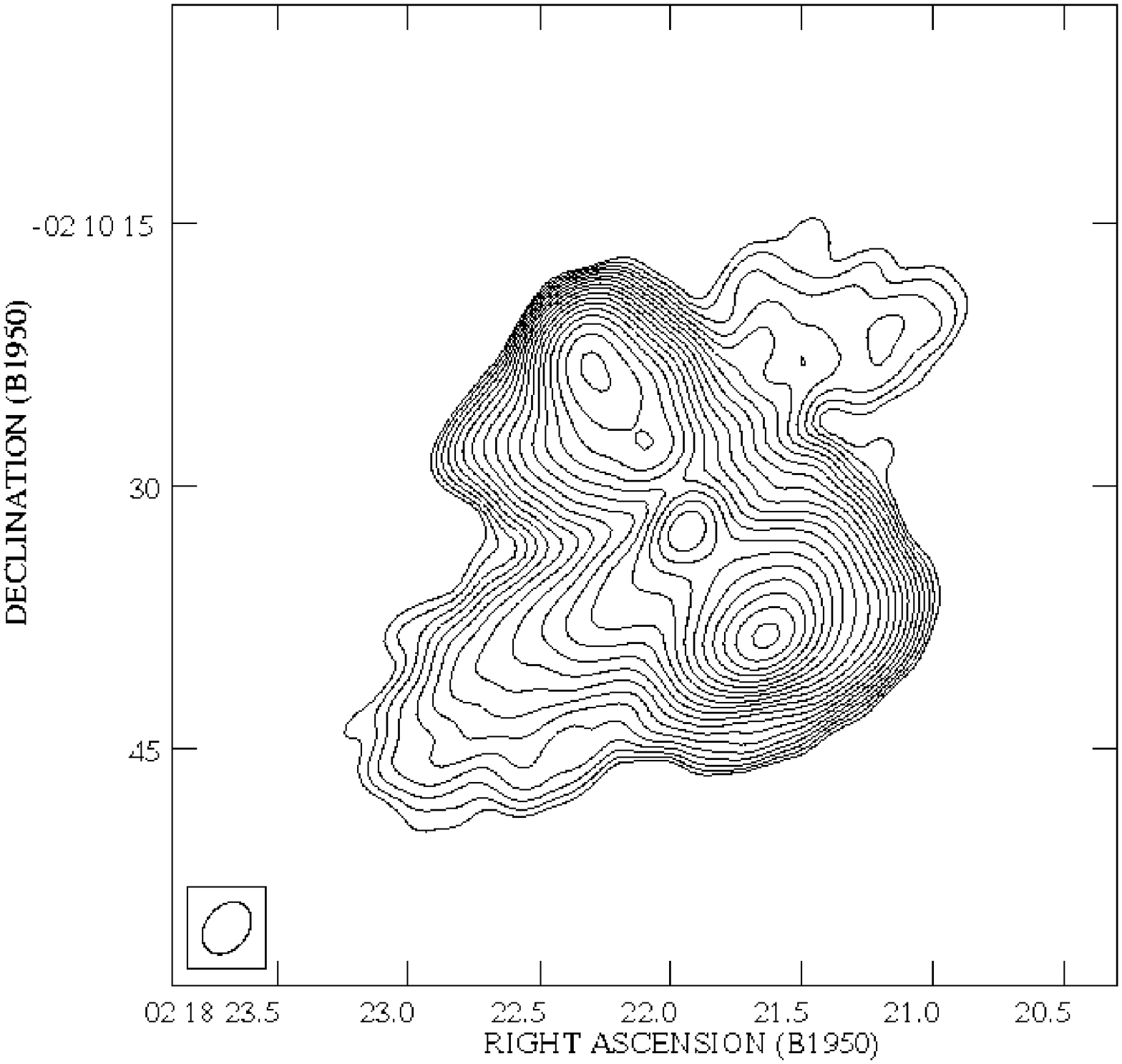}{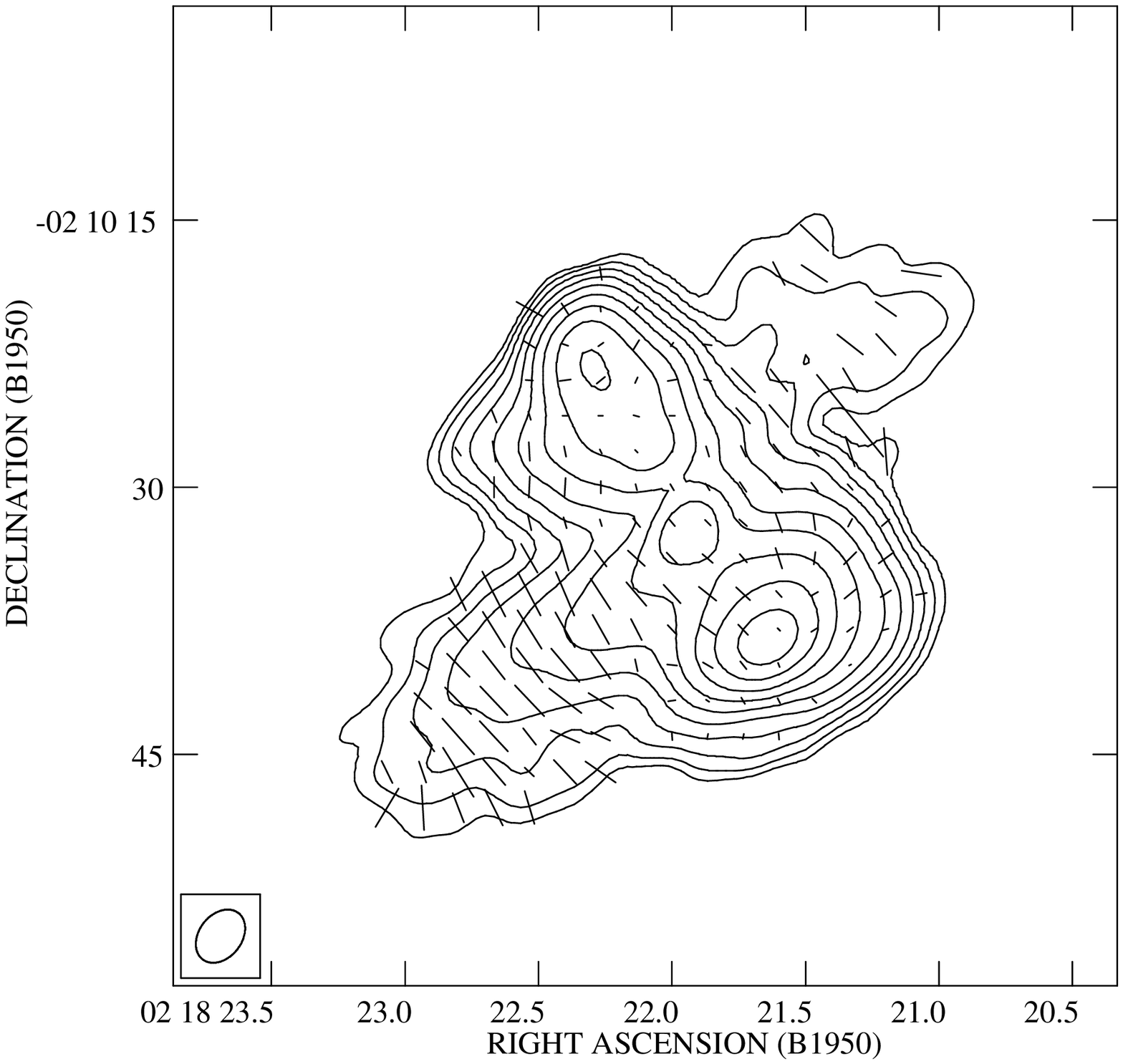}
\figcaption{Left: Total intensity map of 3C\,63 at 8.44 GHz.
The contour levels are $0.1 \times (-\protect\sqrt 2, -1, 1,
\protect\sqrt 2, 2, 2\protect\sqrt 2, \dots)$ mJy beam$^{-1}$. Right:
Polarization map. The contour levels are $0.1 \times (-2, -1, 1, 2, 4,
\dots)$ mJy beam$^{-1}$. A vector of length one arcsecond corresponds
to 20\% polarization. The peak flux occurs in the southern lobe.}
\end{figure}

\begin{figure}
%\epsscale{0.5}
\plotone{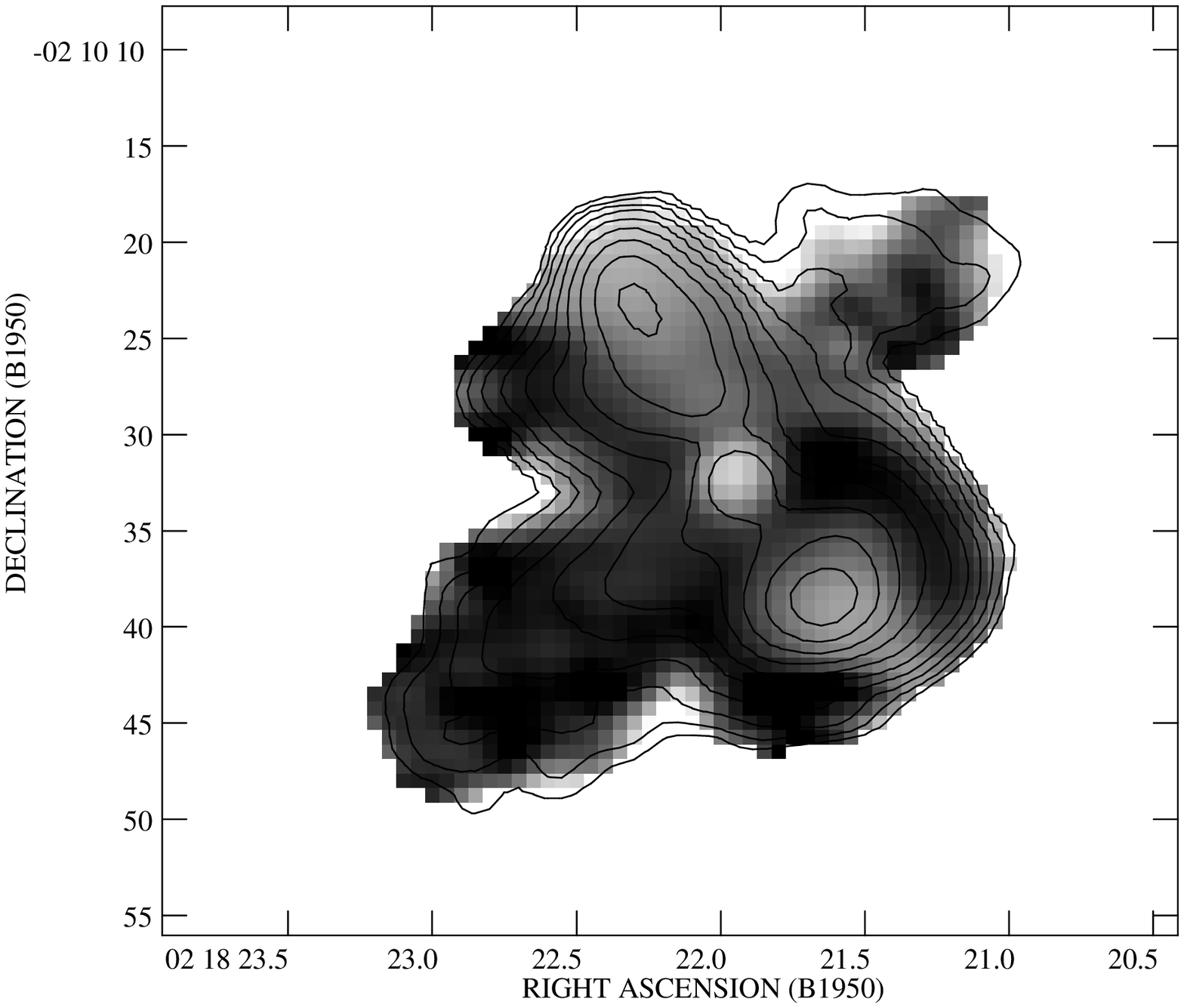}
\figcaption{A 1.4-8.44 GHz spectral index map of 3C\,63 (resolution
$2.83 \times 2.83$ arcsec).  Spectral index values are given by the
greyscale image and range from 0.5 (white) to 1.8 (black).
Superposed are contours of the 8.44-GHz map at this resolution. The
contour levels are $0.1 \times (-2, -1, 1, 2, 4,
\dots)$ mJy beam$^{-1}$.}
%\epsscale{1.0}
\end{figure}

\begin{figure}
\plottwo{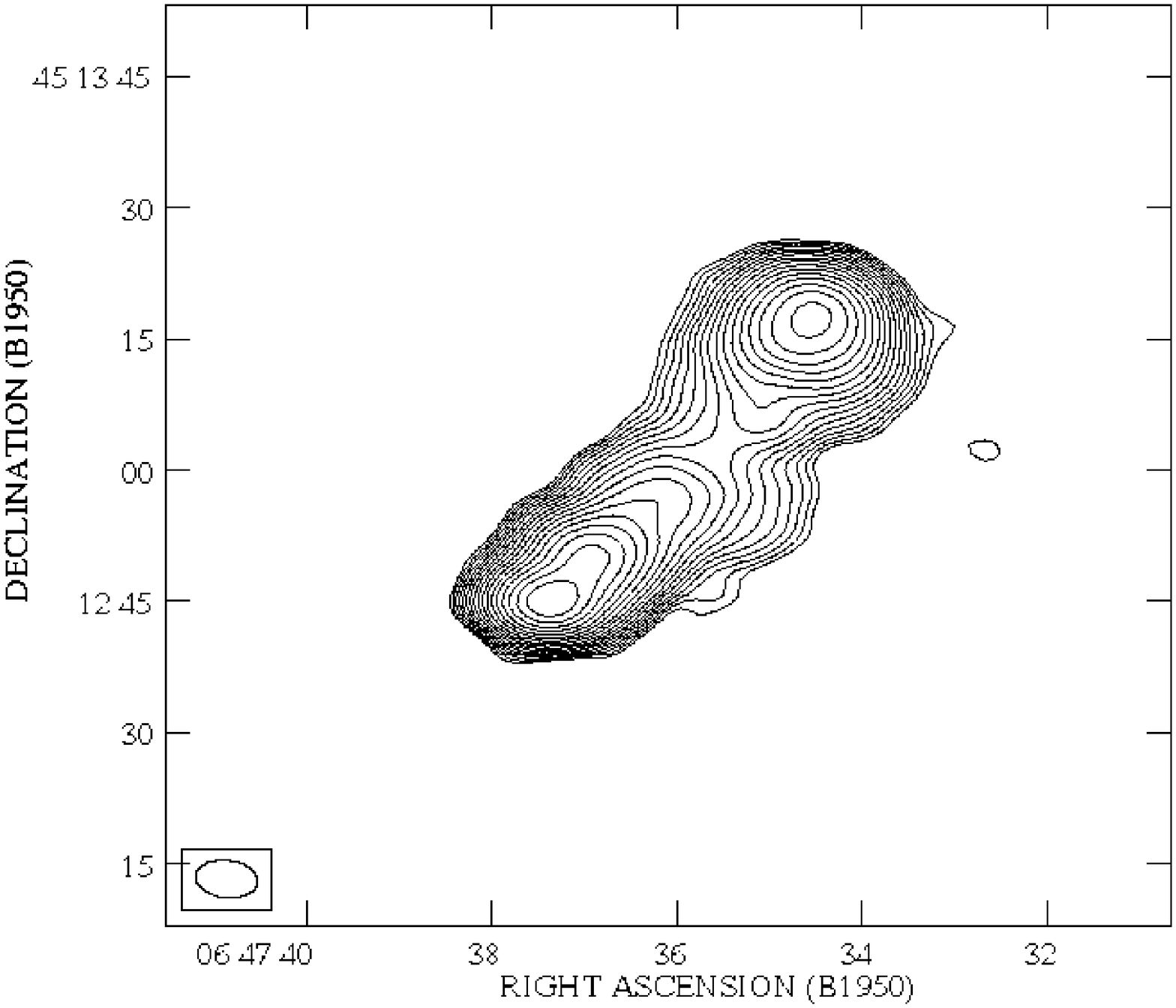}{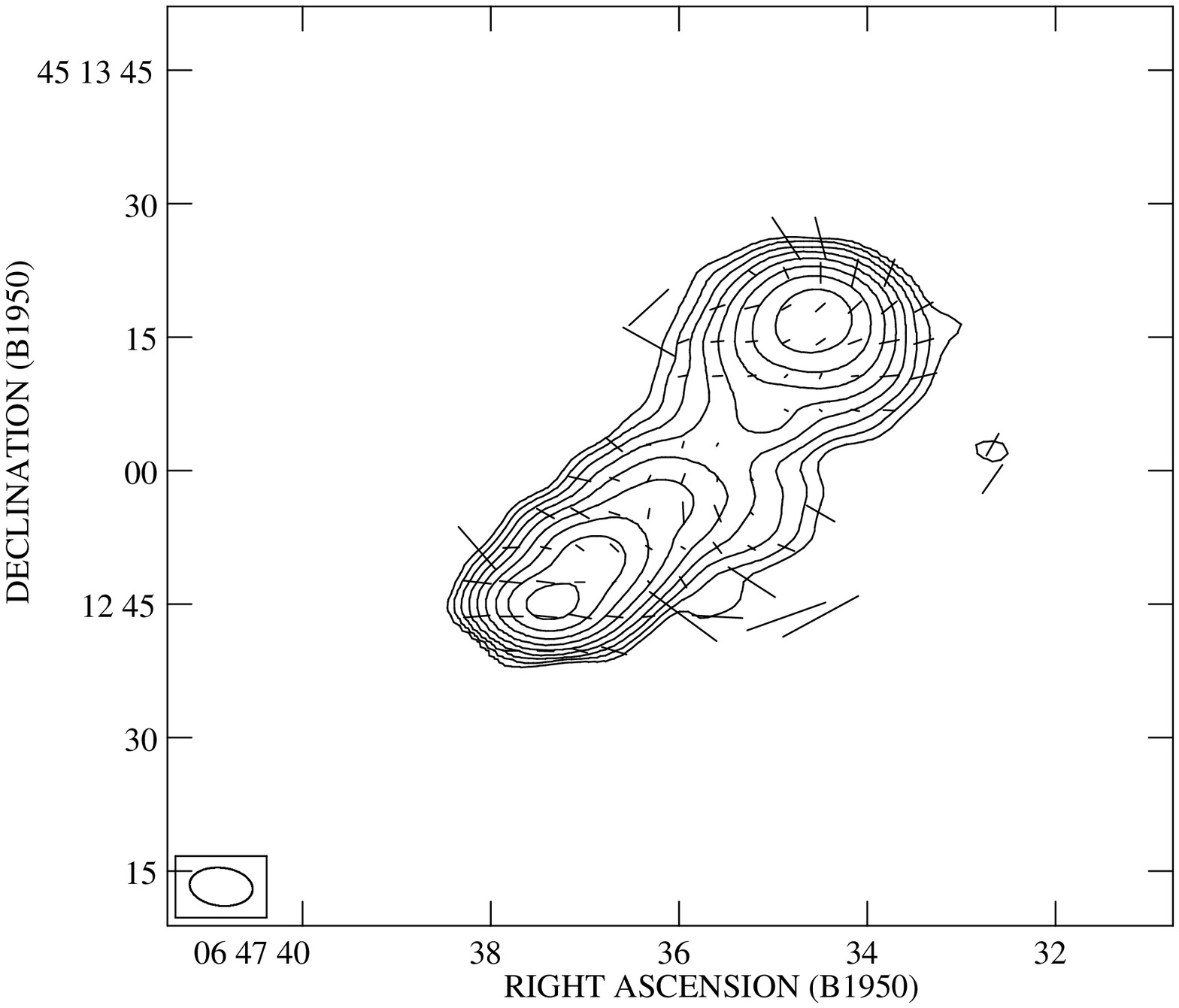}
\figcaption{Left: Total intensity map of 3C\,169.1 at 1.4 GHz.
The contour levels are $0.8 \times (-\protect\sqrt 2, -1, 1,
\protect\sqrt 2, 2, 2\protect\sqrt 2, \dots)$ mJy beam$^{-1}$. Right:
Polarization map. The contour levels are $0.8 \times (-2, -1, 1, 2, 4,
\dots)$ mJy beam$^{-1}$. A vector of length one arcsecond corresponds
to 10\% polarization. The peak flux occurs in the southern lobe.}
\end{figure}

\begin{figure}
\plottwo{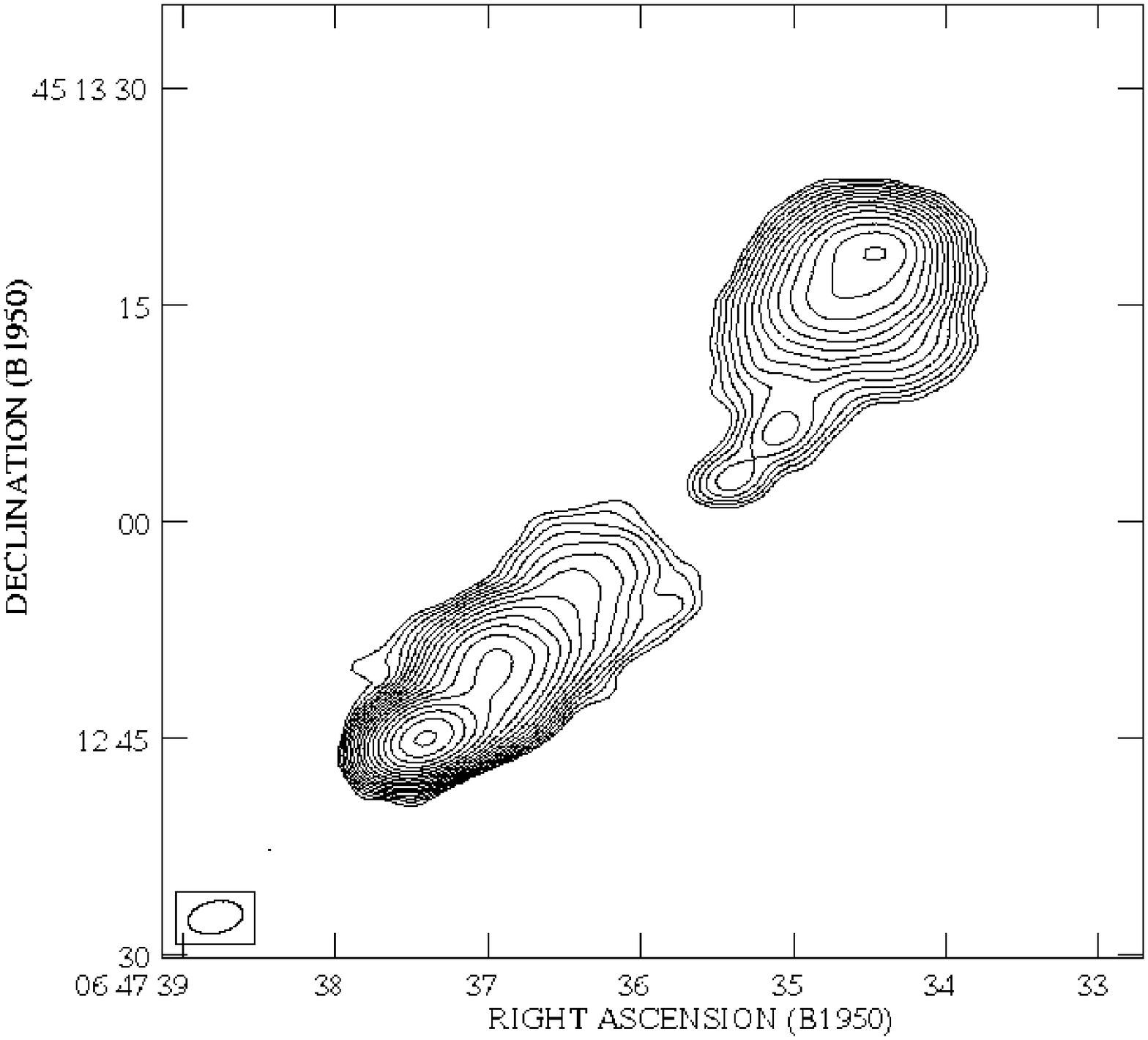}{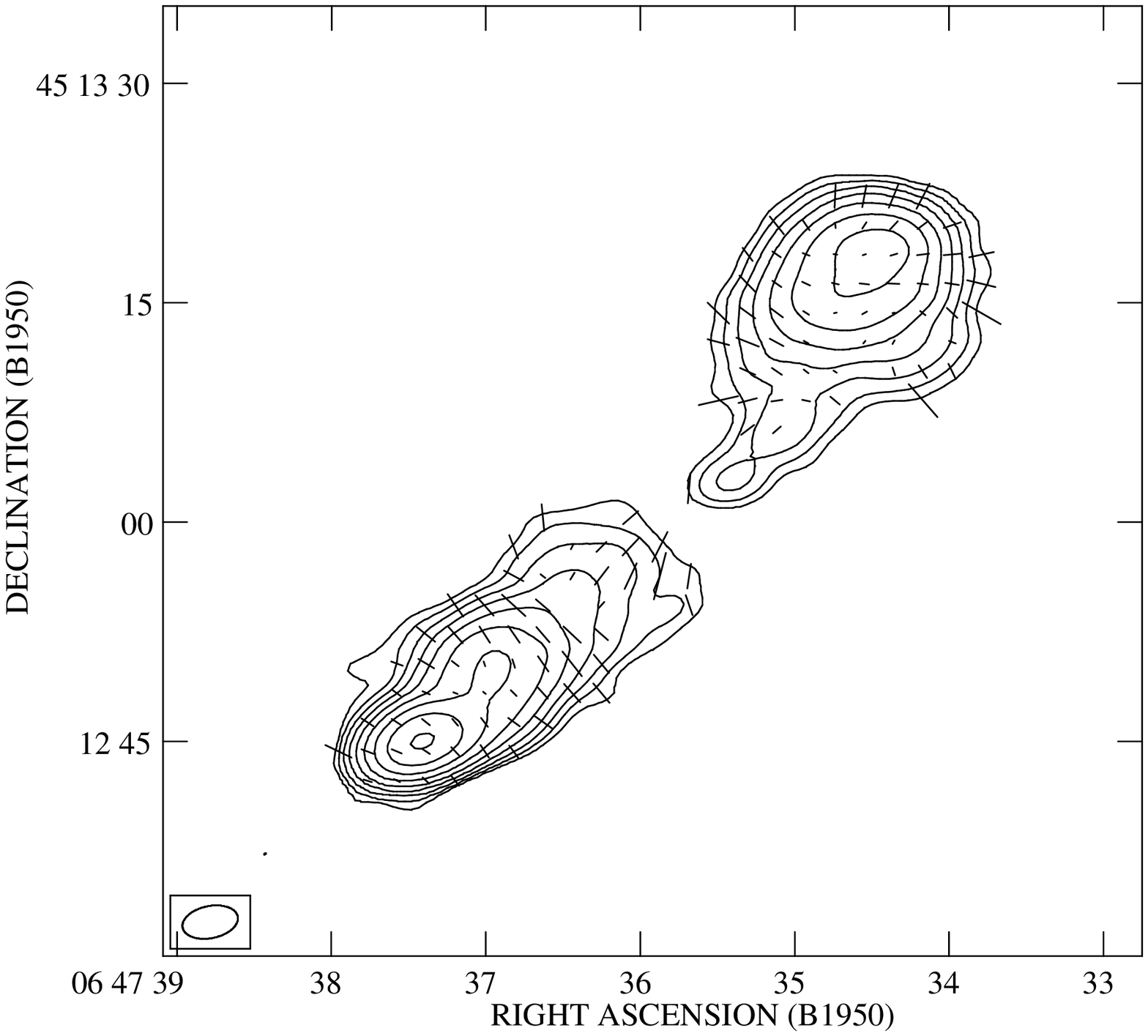}
\figcaption{Left: Total intensity map of 3C\,169.1 at 8.44 GHz.
The contour levels are $0.15 \times (-\protect\sqrt 2, -1, 1,
\protect\sqrt 2, 2, 2\protect\sqrt 2, \dots)$ mJy beam$^{-1}$. Right:
Polarization map. The contour levels are $0.15 \times (-2, -1, 1, 2, 4,
\dots)$ mJy beam$^{-1}$. A vector of length one arcsecond corresponds
to 20\% polarization. The peak flux occurs in the southern lobe.}
\end{figure}

\begin{figure}
%\epsscale{0.5}
\plotone{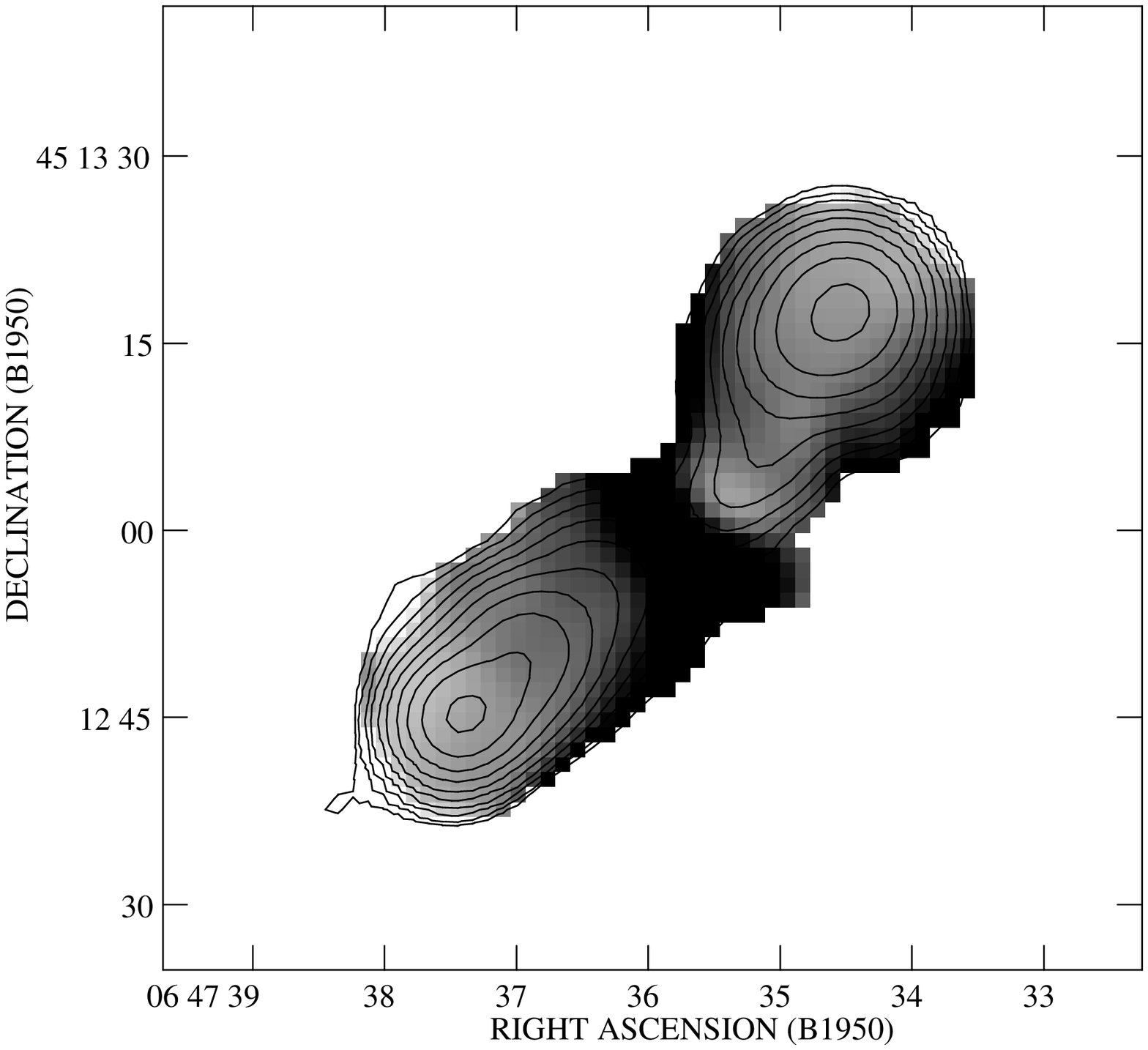}
\figcaption{A 1.4-8.44 GHz spectral index map of 3C\,169.1 (resolution
$5.40 \times 5.40$ arcsec).  Spectral index values are given by the
greyscale image and range from 0.5 (white) to 1.8 (black).
Superposed are contours of the 8.44-GHz map at this resolution. The
contour levels are $0.1 \times (-2, -1, 1, 2, 4,
\dots)$ mJy beam$^{-1}$.}
%\epsscale{1.0}
\end{figure}

\begin{figure}
\plottwo{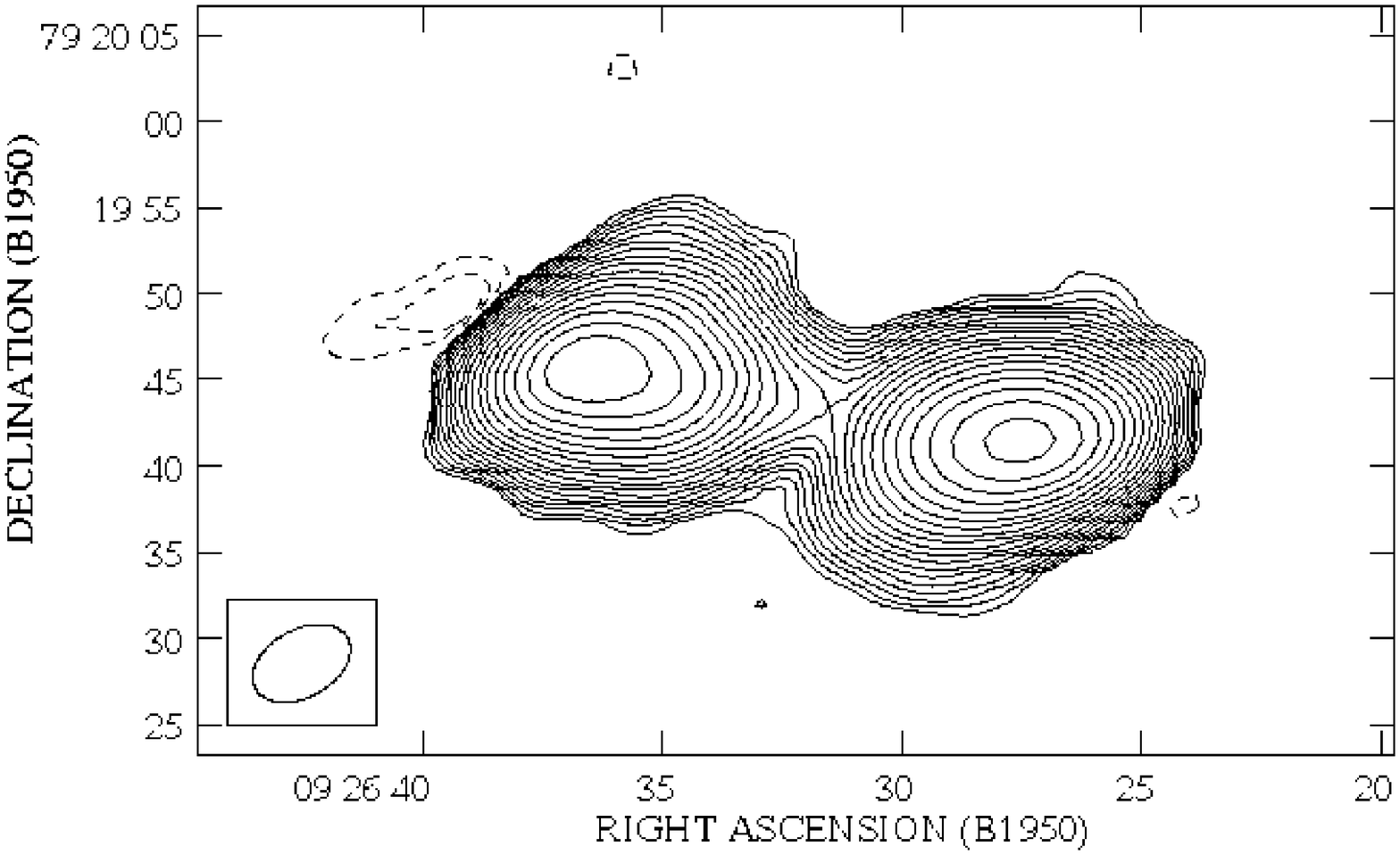}{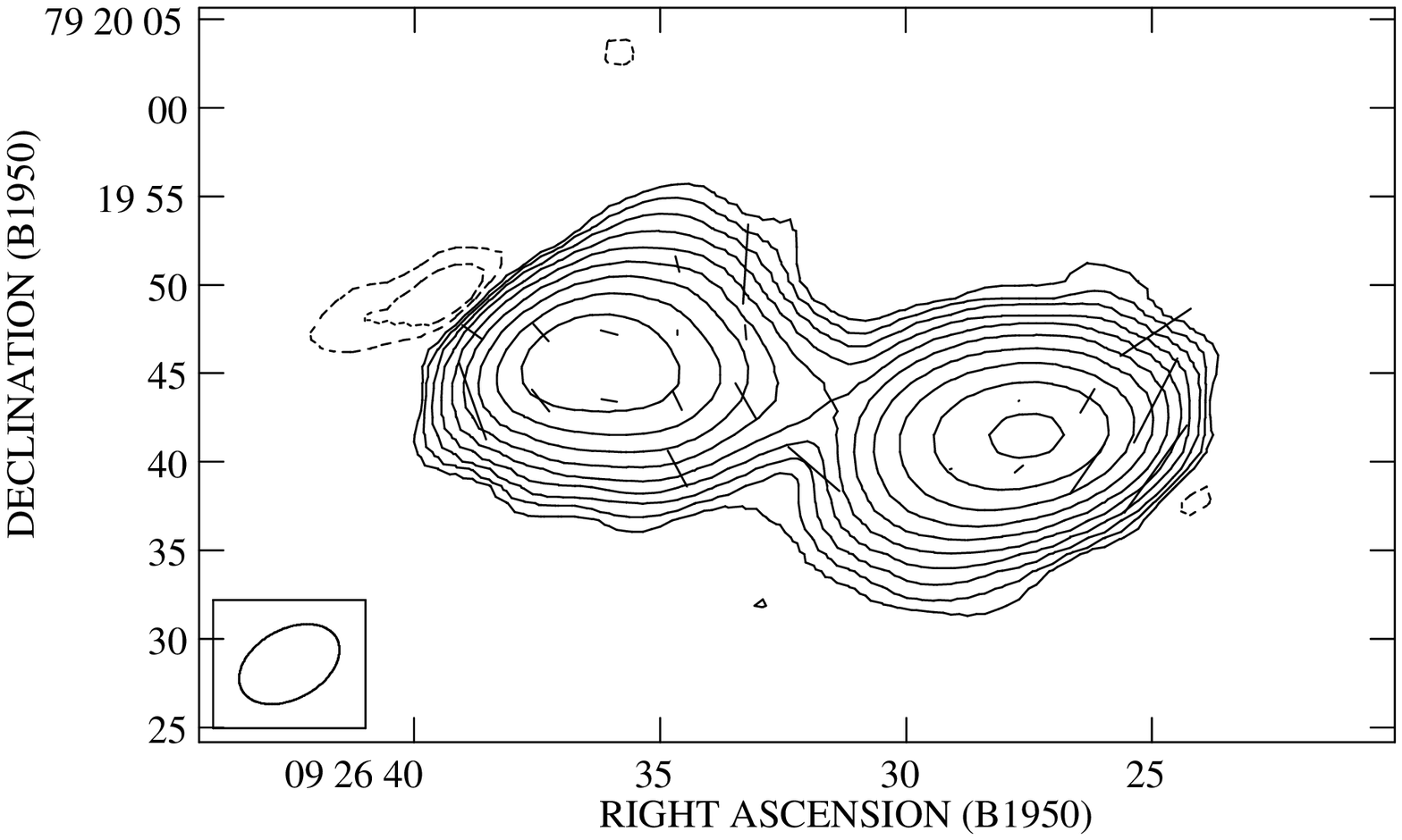}
\figcaption{Left: Total intensity map of 3C\,220.1 at 1.4 GHz.
The contour levels are $1.0 \times (-\protect\sqrt 2, -1, 1,
\protect\sqrt 2, 2, 2\protect\sqrt 2, \dots)$ mJy beam$^{-1}$. Right:
Polarization map. The contour levels are $1.0 \times (-2, -1, 1, 2, 4,
\dots)$ mJy beam$^{-1}$. A vector of length one arcsecond corresponds
to 1\% polarization. The peak flux occurs in the western lobe.}
\end{figure}

\begin{figure}
\plottwo{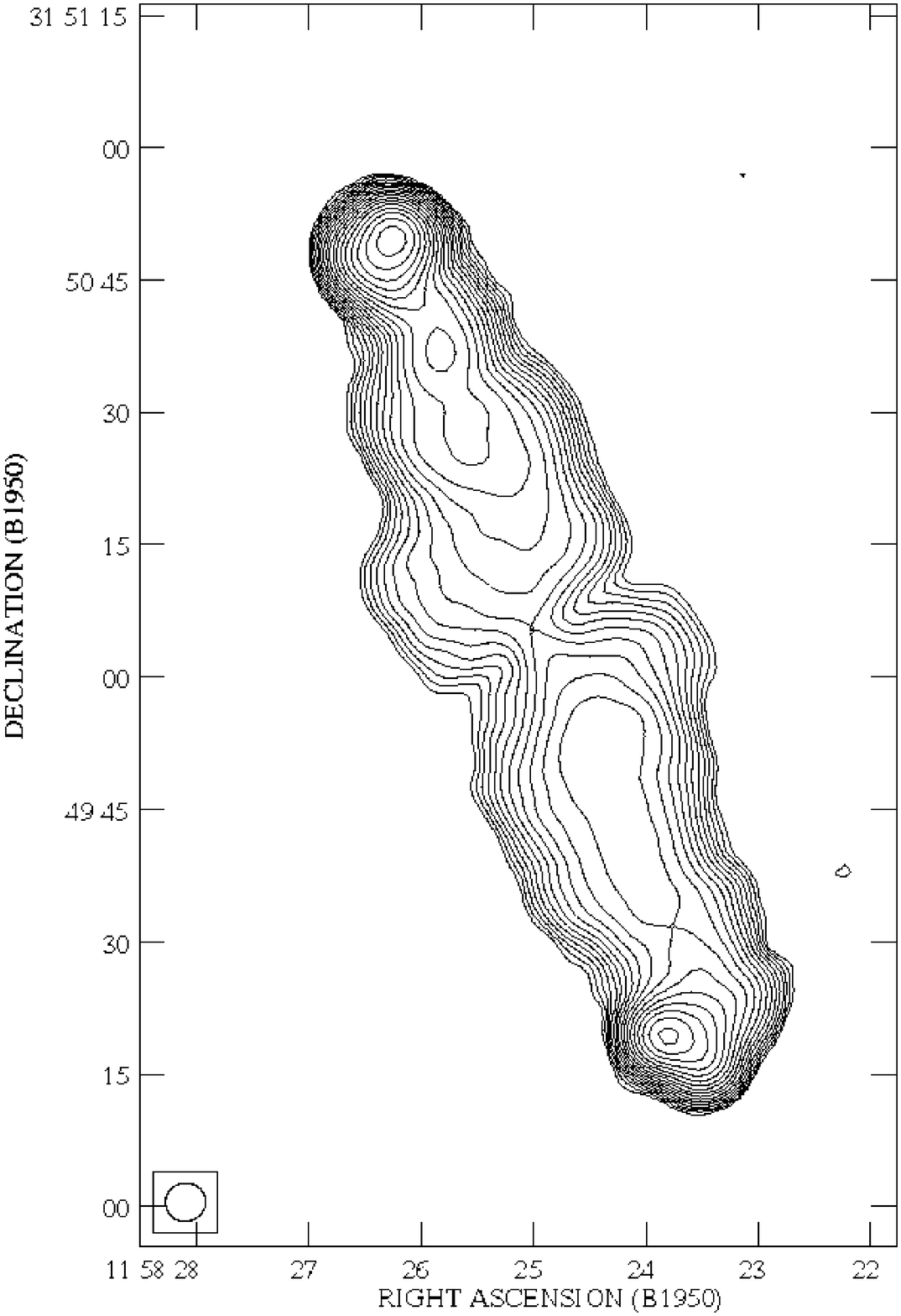}{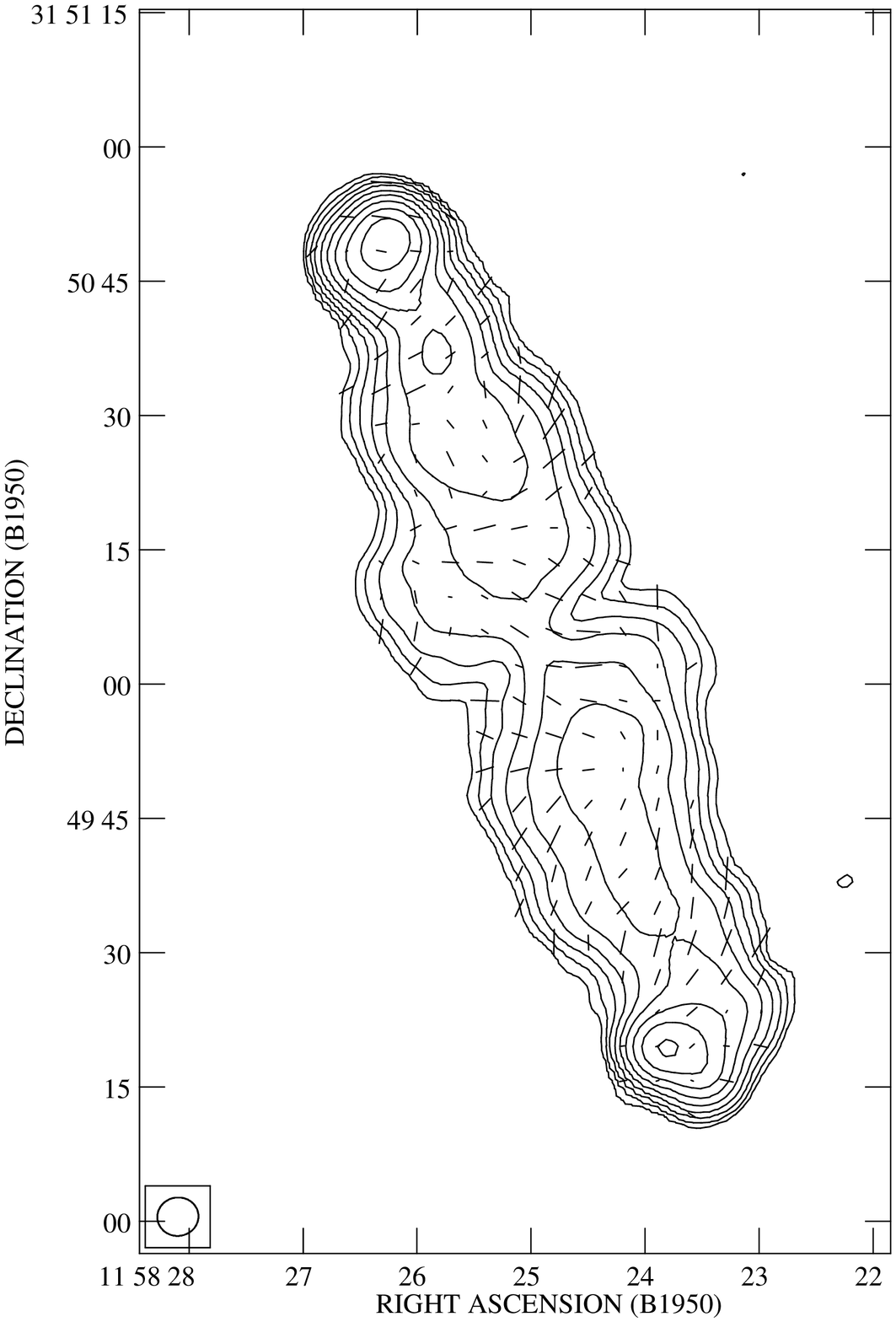}
\figcaption{Left: Total intensity map of 3C\,268.2 at 1.4 GHz.
The contour levels are $0.4 \times (-\protect\sqrt 2, -1, 1,
\protect\sqrt 2, 2, 2\protect\sqrt 2, \dots)$ mJy beam$^{-1}$. Right:
Polarization map. The contour levels are $0.4 \times (-2, -1, 1, 2, 4,
\dots)$ mJy beam$^{-1}$. A vector of length one arcsecond corresponds
to 20\% polarization. The peak flux occurs in the northern lobe.}
\end{figure}

\begin{figure}
\plottwo{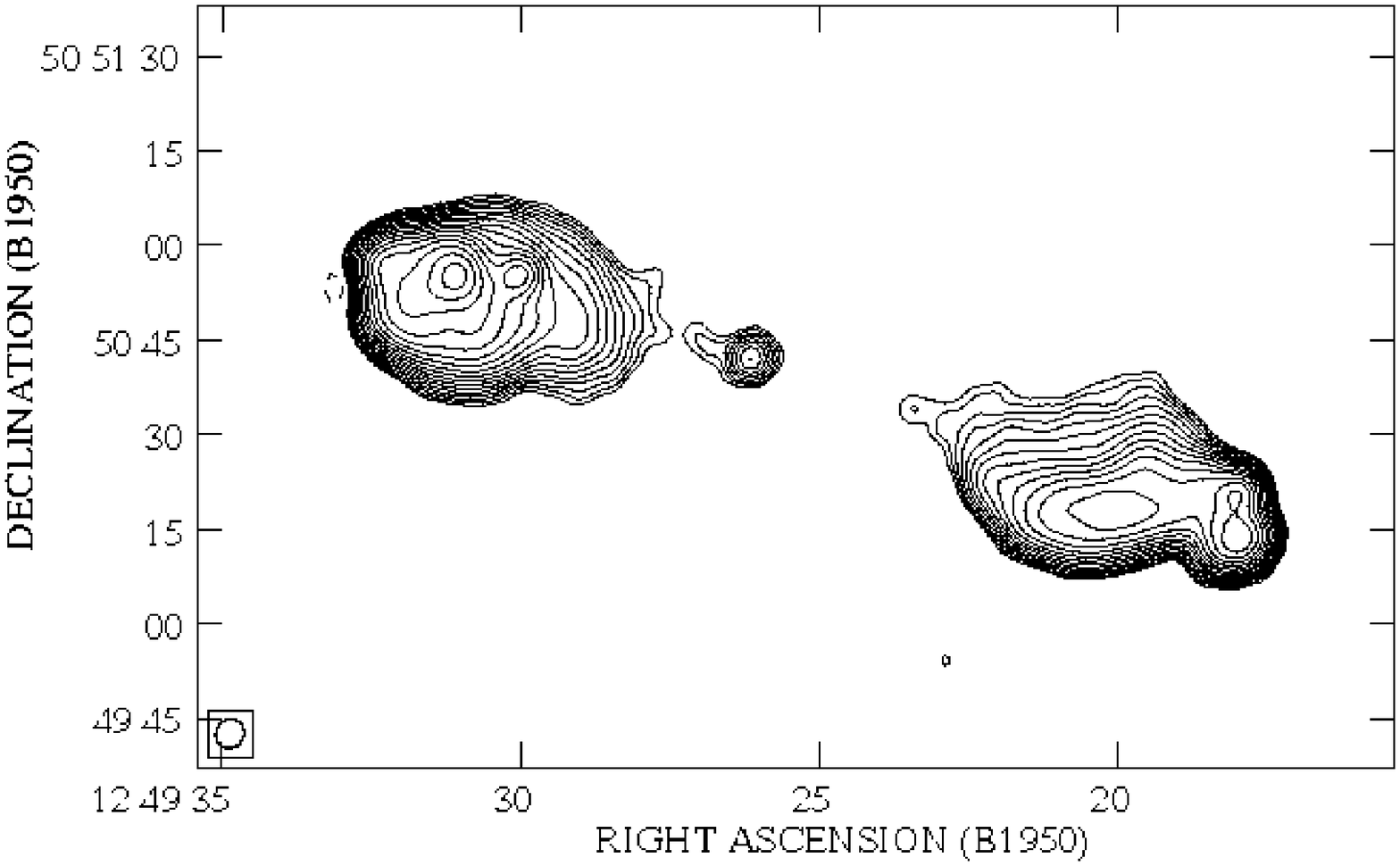}{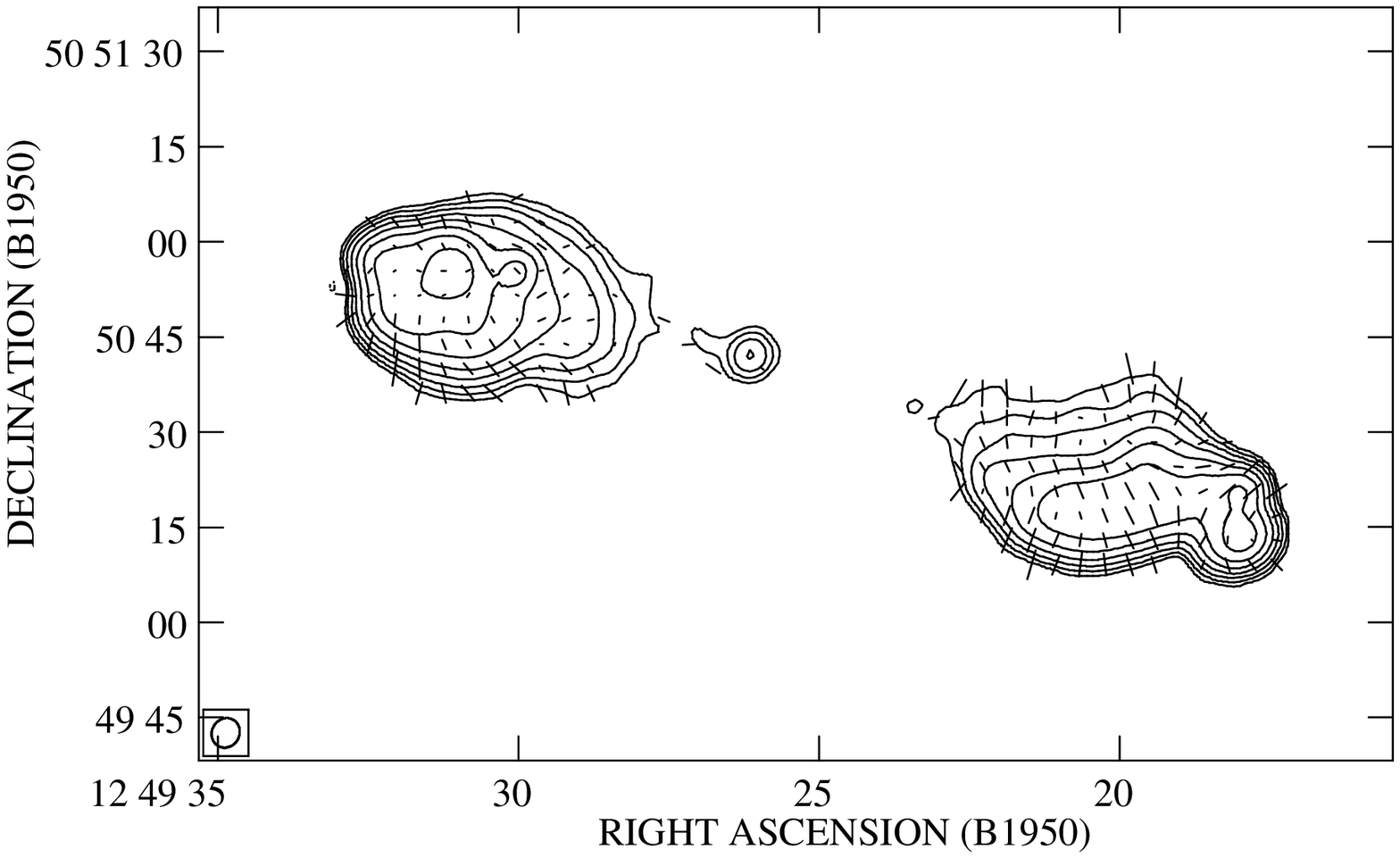}
\figcaption{Left: Total intensity map of 3C\,277 at 1.4 GHz.
The contour levels are $0.3 \times (-\protect\sqrt 2, -1, 1,
\protect\sqrt 2, 2, 2\protect\sqrt 2, \dots)$ mJy beam$^{-1}$. Right:
Polarization map. The contour levels are $0.3 \times (-2, -1, 1, 2, 4,
\dots)$ mJy beam$^{-1}$. A vector of length one arcsecond corresponds
to 20\% polarization. The peak flux occurs in the eastern lobe.}
\end{figure}

\begin{figure}
\plottwo{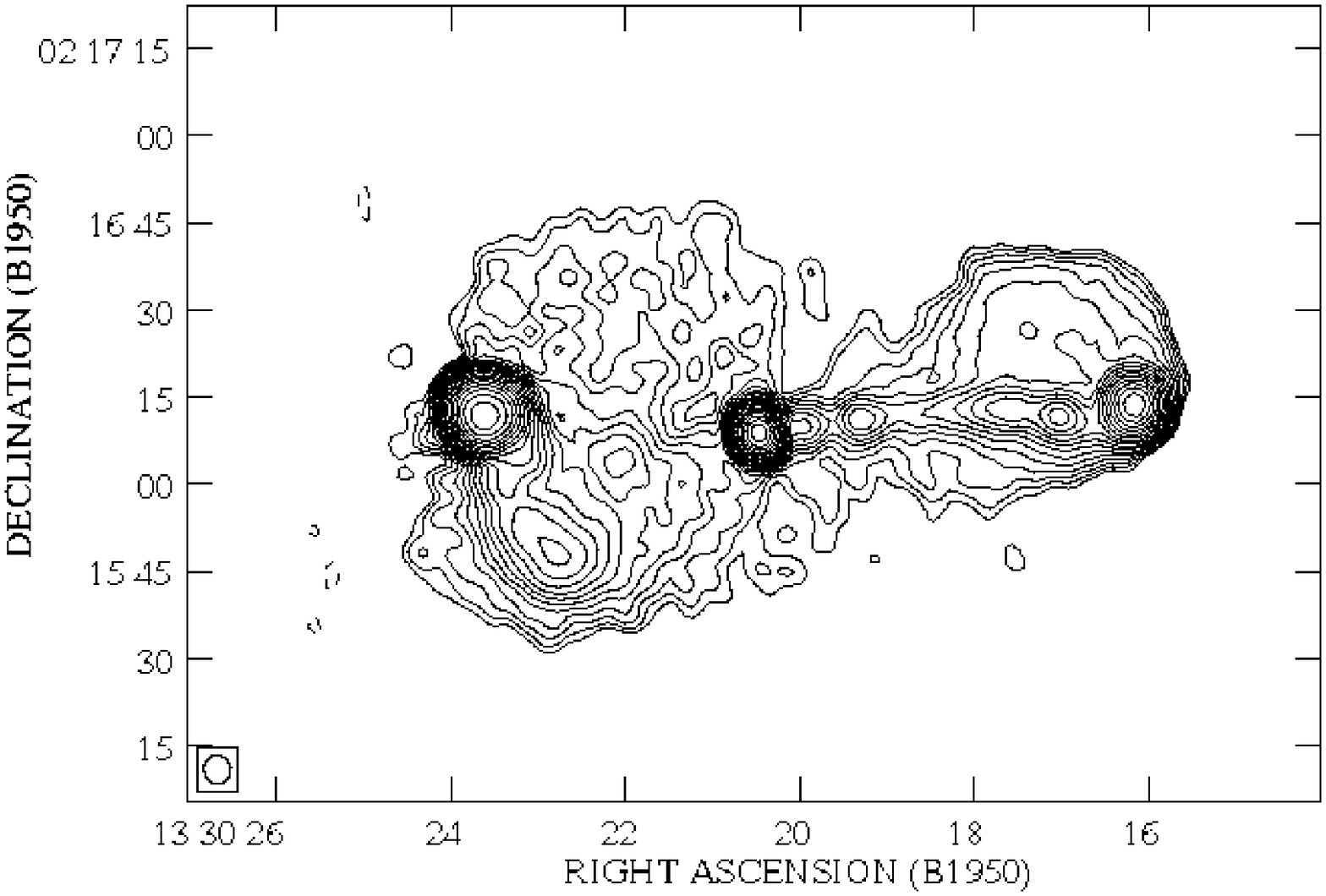}{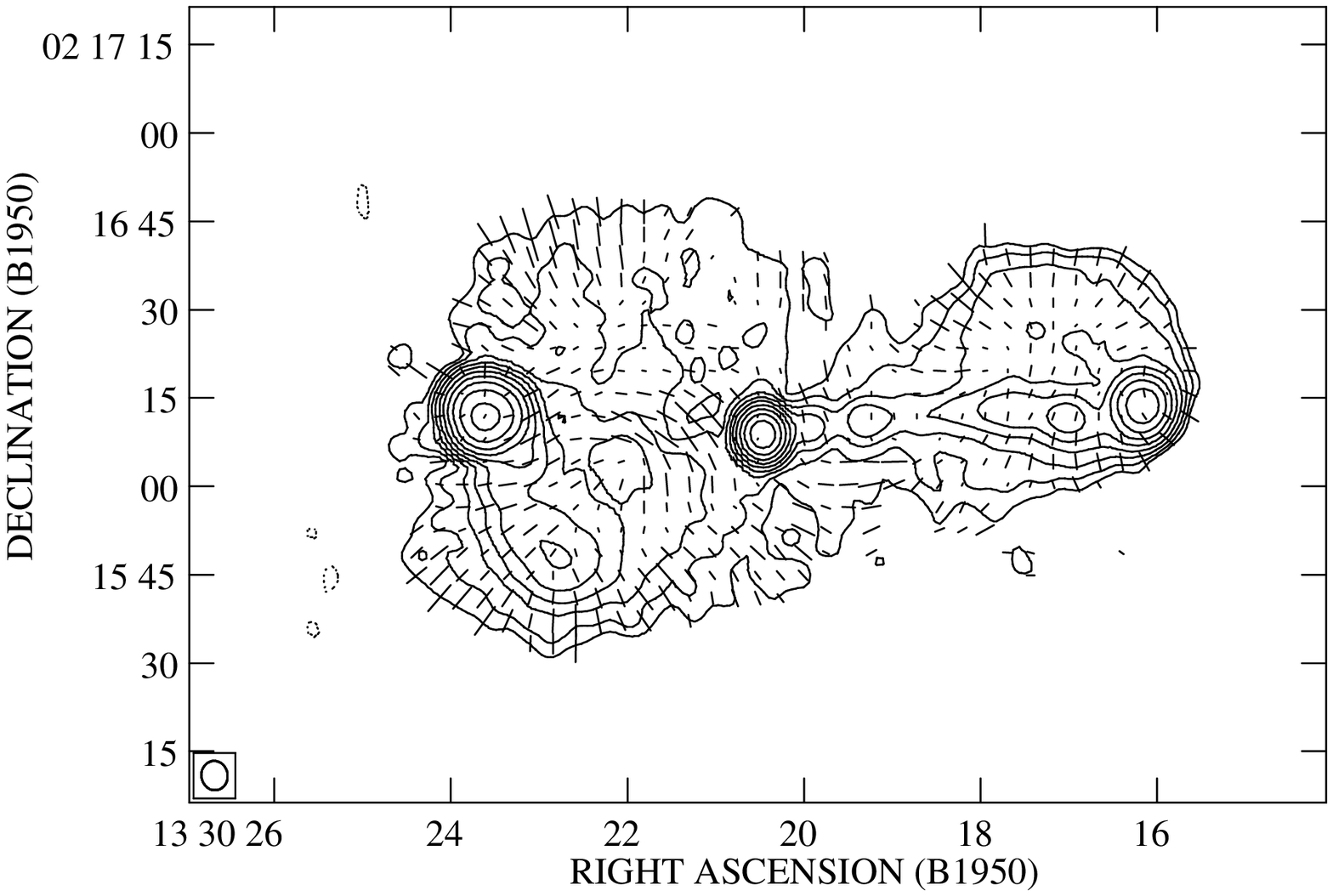}
\figcaption{Left: Total intensity map of 3C\,287.1 at 1.4 GHz.
The contour levels are $0.7 \times (-\protect\sqrt 2, -1, 1,
\protect\sqrt 2, 2, 2\protect\sqrt 2, \dots)$ mJy beam$^{-1}$. Right:
Polarization map. The contour levels are $0.7 \times (-2, -1, 1, 2, 4,
\dots)$ mJy beam$^{-1}$. A vector of length one arcsecond corresponds
to 20\% polarization. The peak flux occurs in the hotspot in the eastern lobe.}
\end{figure}

\begin{figure}
\plottwo{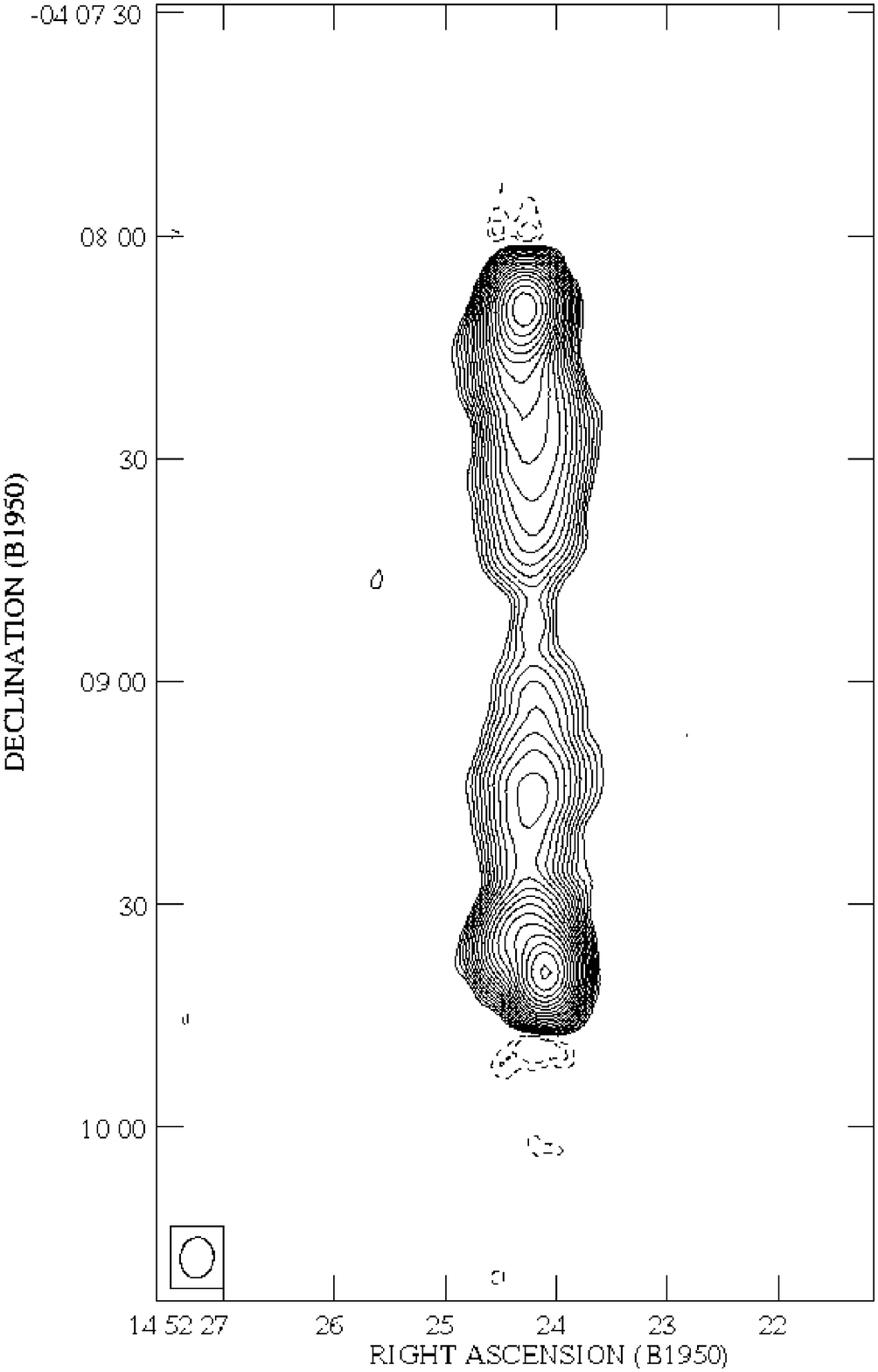}{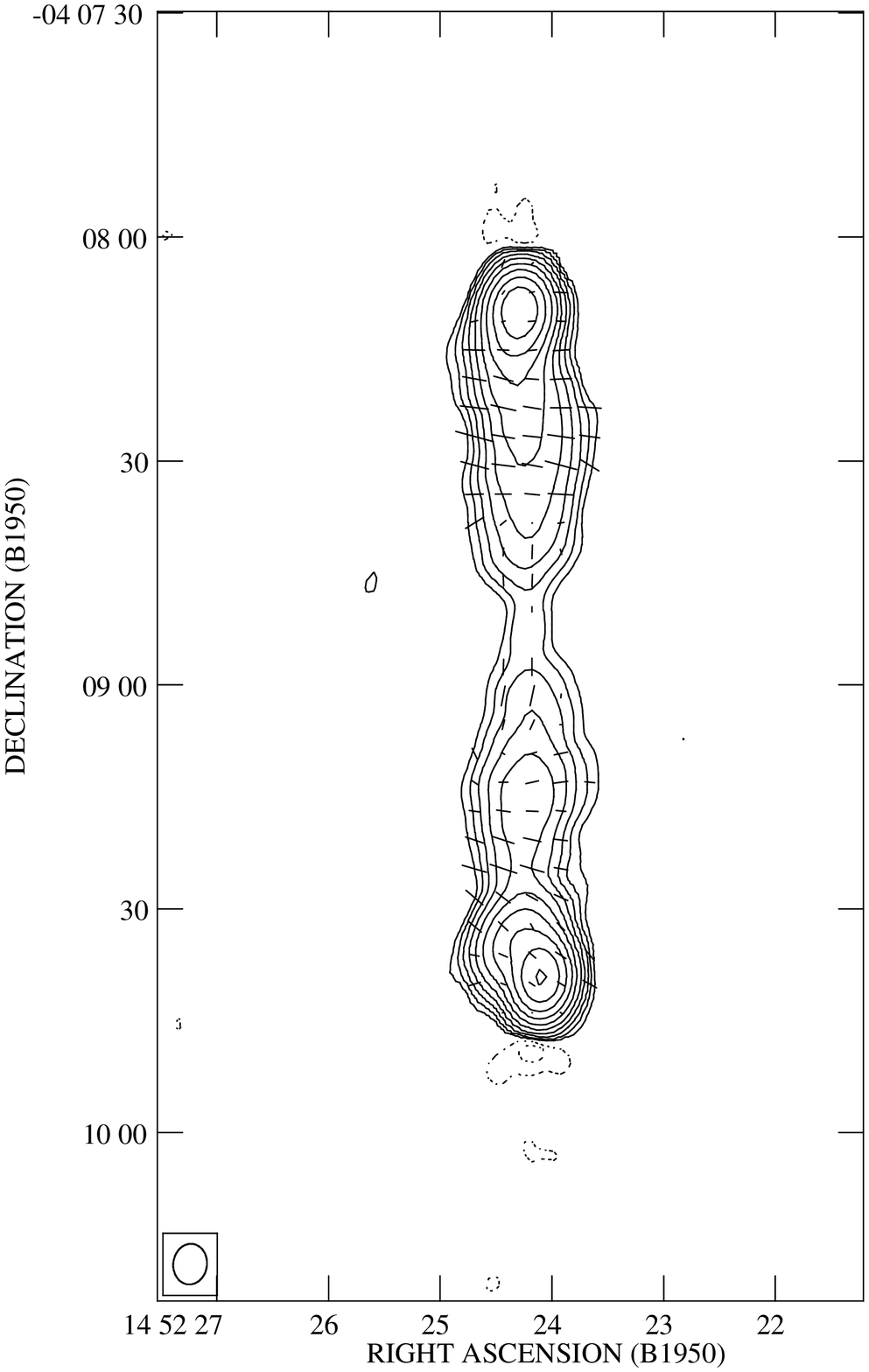}
\figcaption{Left: Total intensity map of 3C\,306.1 at 1.4 GHz.
The contour levels are $1.0 \times (-\protect\sqrt 2, -1, 1,
\protect\sqrt 2, 2, 2\protect\sqrt 2, \dots)$ mJy beam$^{-1}$. Right:
Polarization map. The contour levels are $1.0 \times (-2, -1, 1, 2, 4,
\dots)$ mJy beam$^{-1}$. A vector of length one arcsecond corresponds
to 20\% polarization. The peak flux occurs in the southern lobe.}
\end{figure}

\begin{figure}
\plottwo{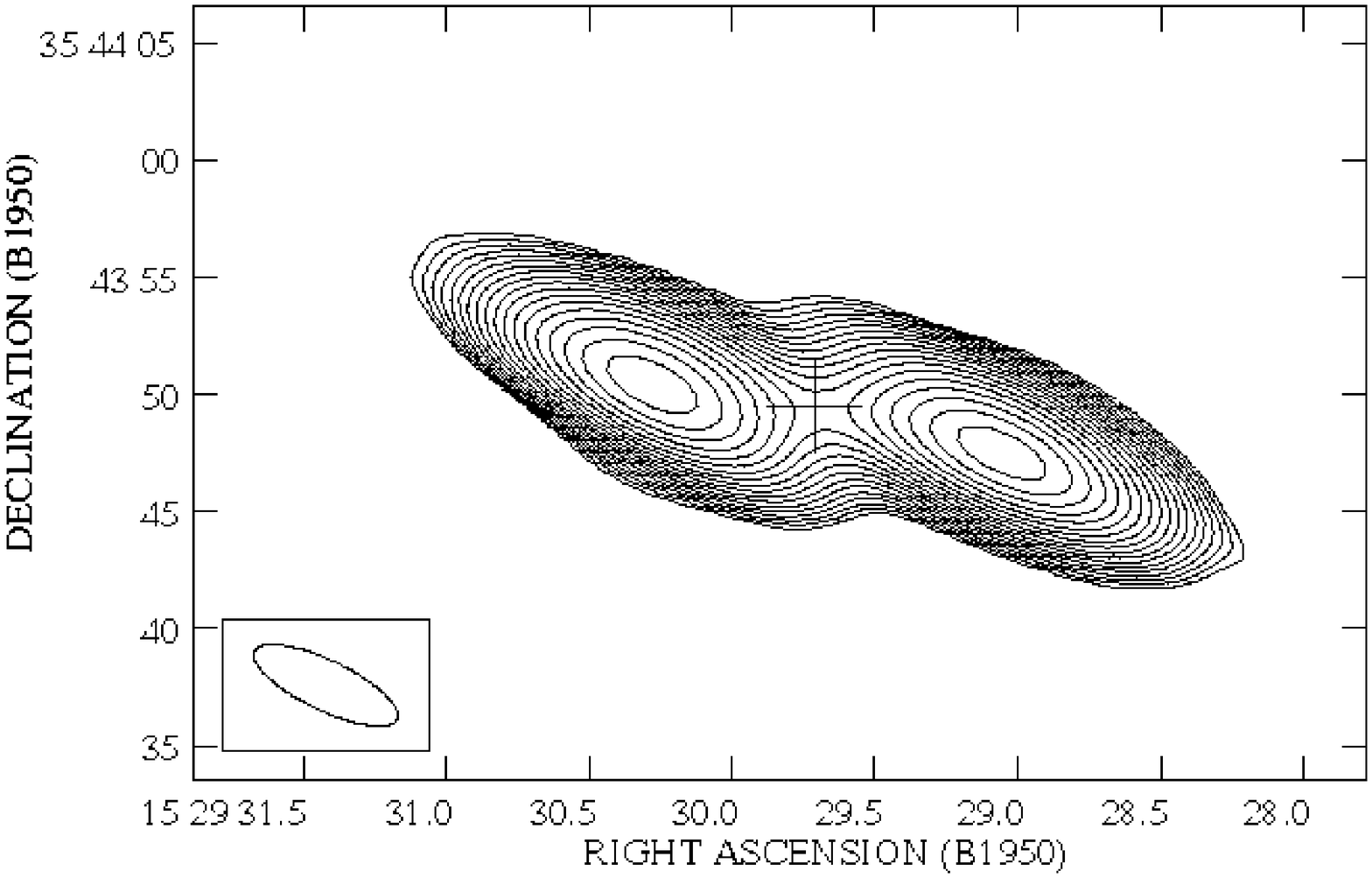}{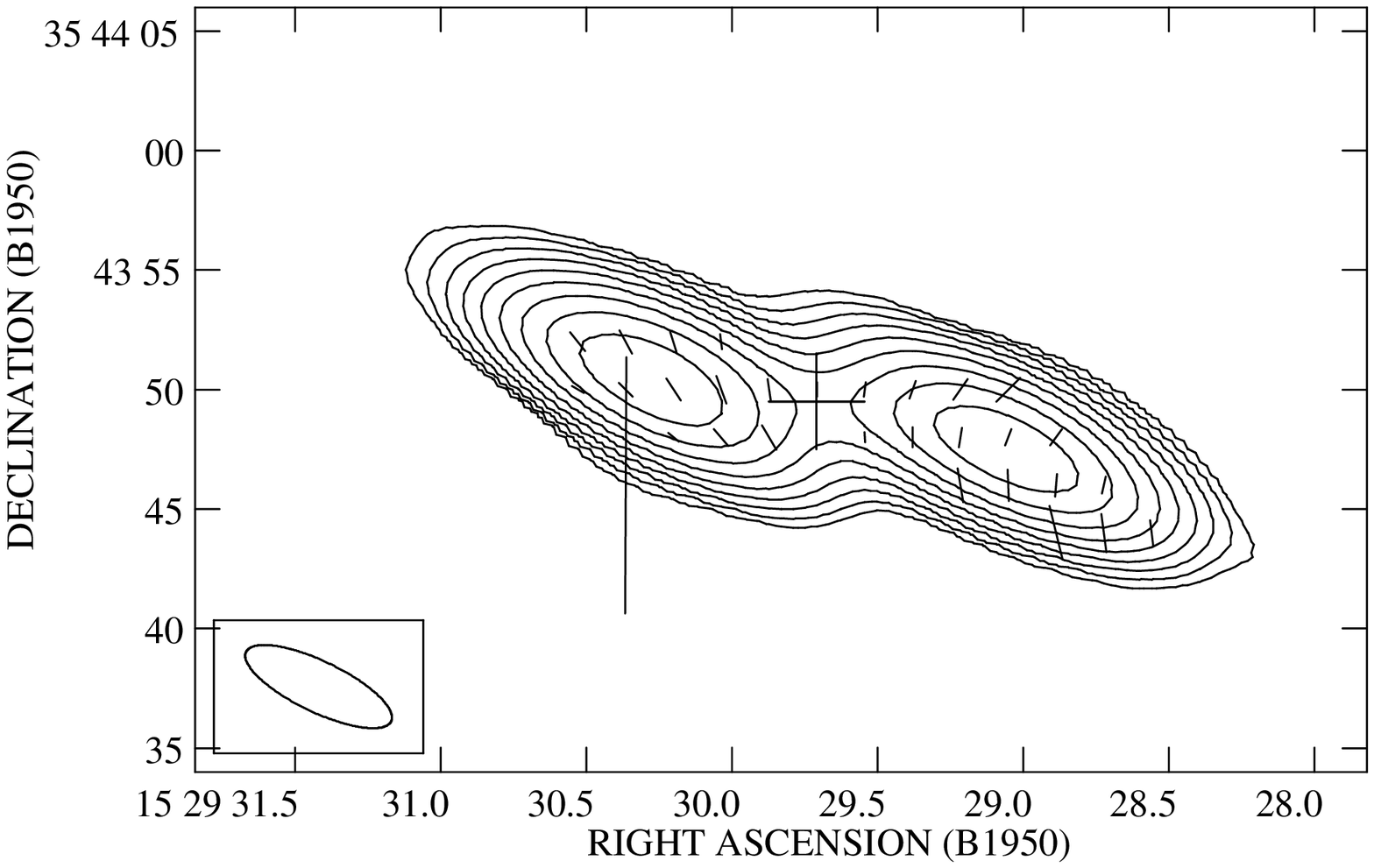}
\figcaption{Left: Total intensity map of 3C\,320 at 8.44 GHz.
The contour levels are $0.1 \times (-\protect\sqrt 2, -1, 1,
\protect\sqrt 2, 2, 2\protect\sqrt 2, \dots)$ mJy beam$^{-1}$. Right:
Polarization map. The contour levels are $0.1 \times (-2, -1, 1, 2, 4,
\dots)$ mJy beam$^{-1}$. A vector of length one arcsecond corresponds
to 5\% polarization. A cross marks the position of the probable core
discussed in the text. The peak flux occurs in the eastern lobe.}
\end{figure}

\begin{figure}
\plottwo{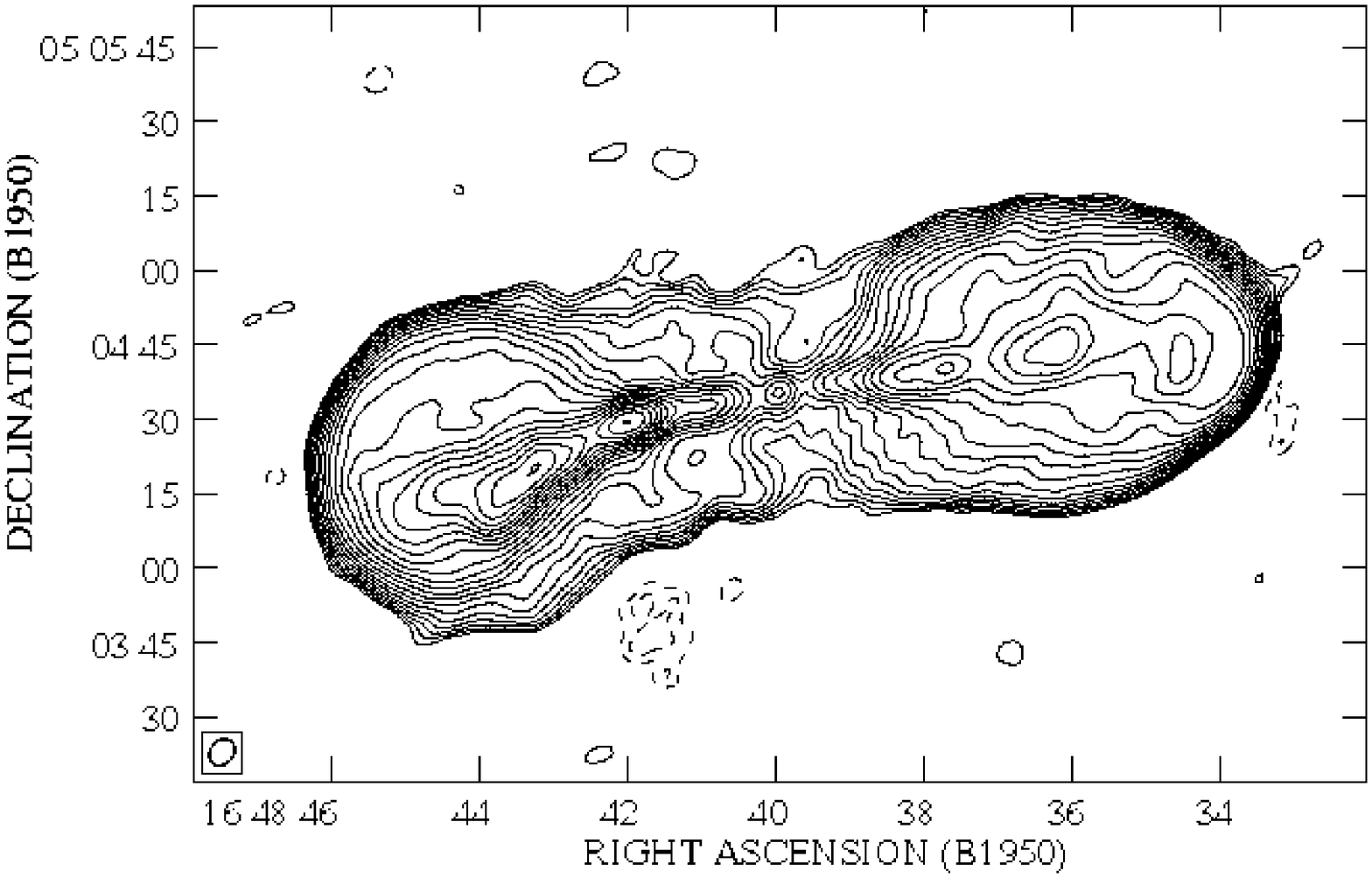}{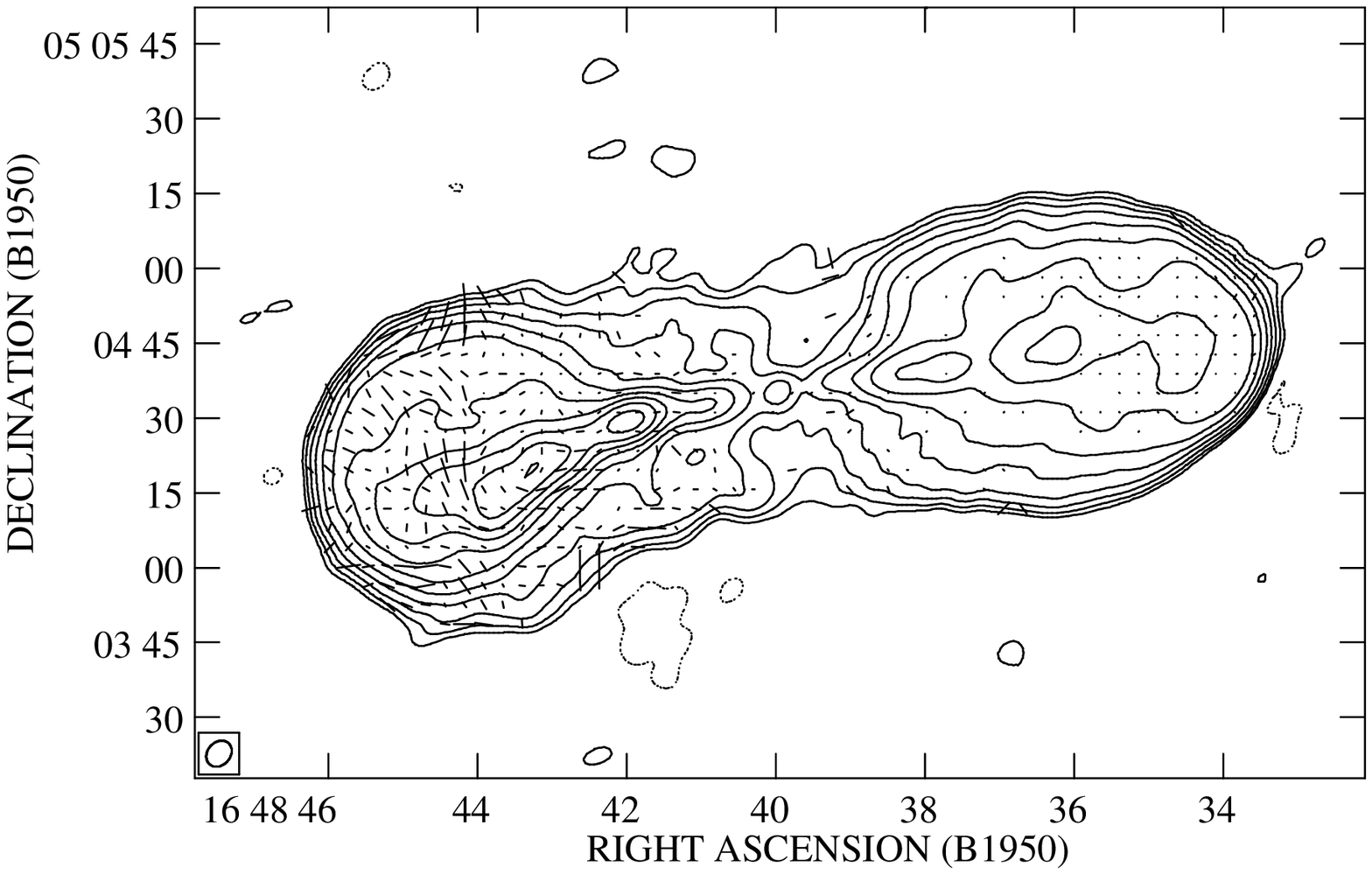}
\figcaption{Left: Total intensity map of 3C\,348 at 1.4 GHz.
The contour levels are $2.0 \times (-\protect\sqrt 2, -1, 1,
\protect\sqrt 2, 2, 2\protect\sqrt 2, \dots)$ mJy beam$^{-1}$. Right:
Polarization map. The contour levels are $2.0 \times (-2, -1, 1, 2, 4,
\dots)$ mJy beam$^{-1}$. A vector of length one arcsecond corresponds
to 10\% polarization. The peak flux occurs in the jet in the eastern lobe.}
\end{figure}

\begin{figure}
\plottwo{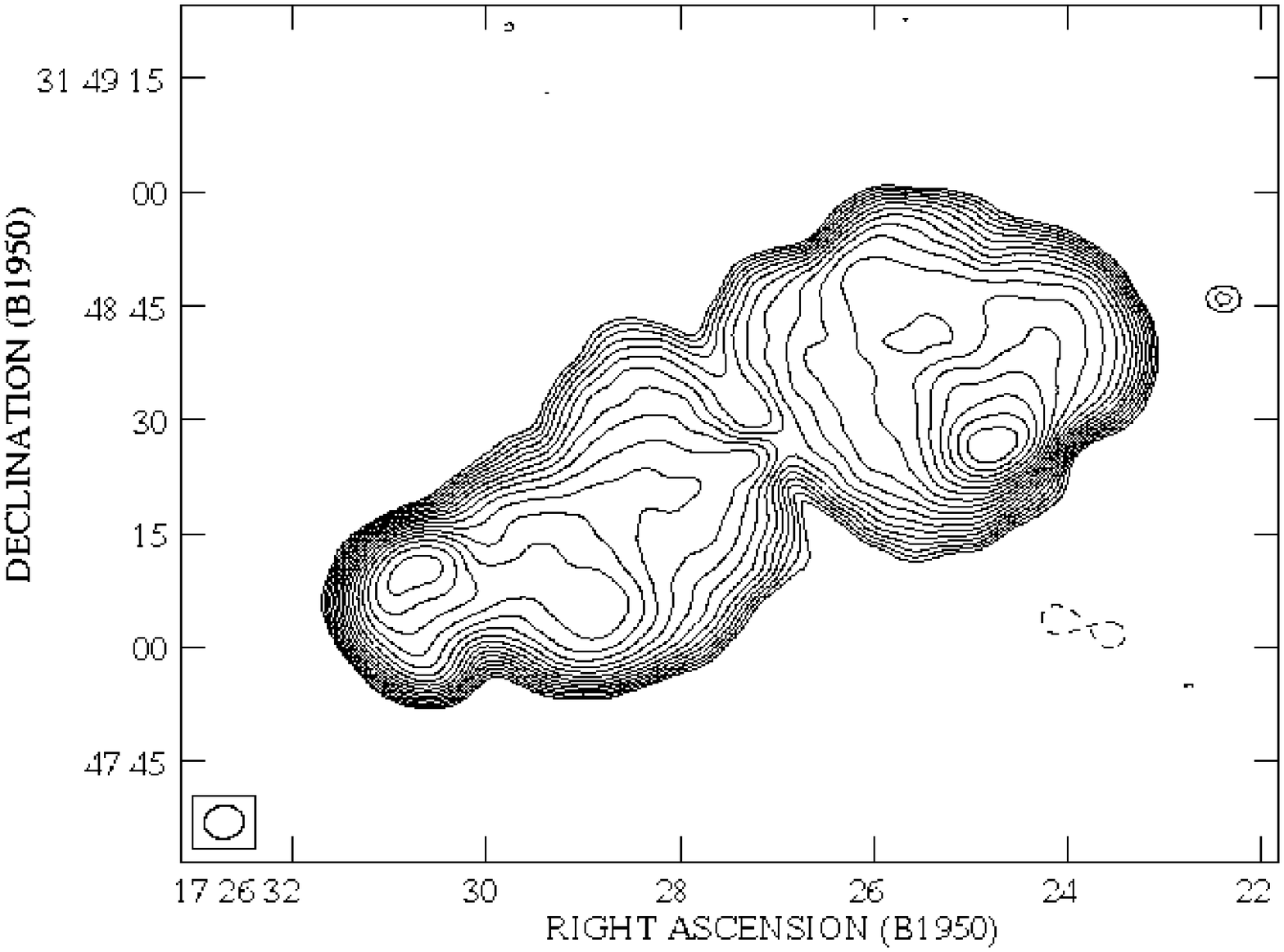}{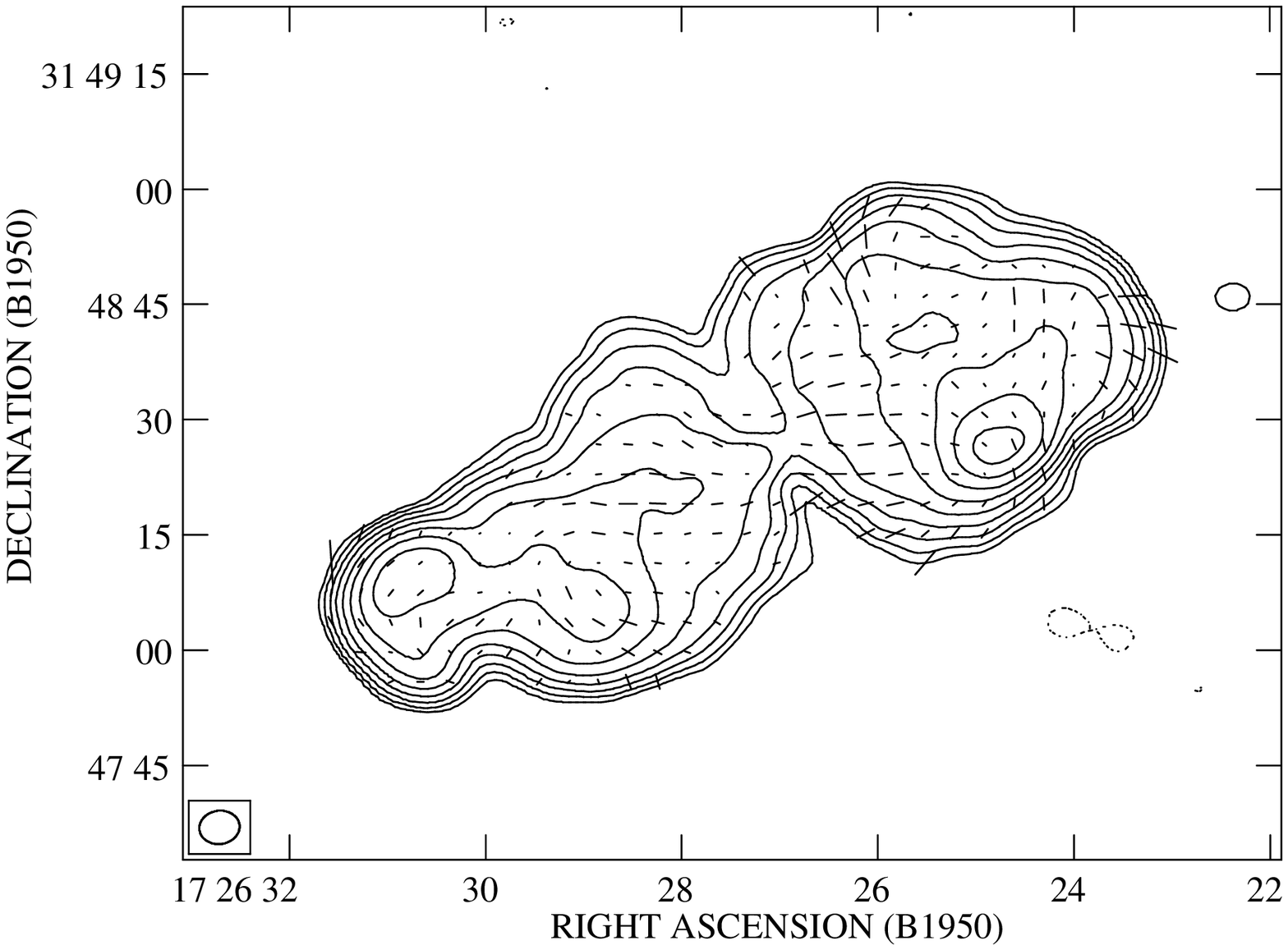}
\figcaption{Left: Total intensity map of 3C\,357 at 1.4 GHz.
The contour levels are $0.5 \times (-\protect\sqrt 2, -1, 1,
\protect\sqrt 2, 2, 2\protect\sqrt 2, \dots)$ mJy beam$^{-1}$. Right:
Polarization map. The contour levels are $0.5 \times (-2, -1, 1, 2, 4,
\dots)$ mJy beam$^{-1}$. A vector of length one arcsecond corresponds
to 20\% polarization. The peak flux occurs in the western lobe.}
\end{figure}

\begin{figure}
\plotone{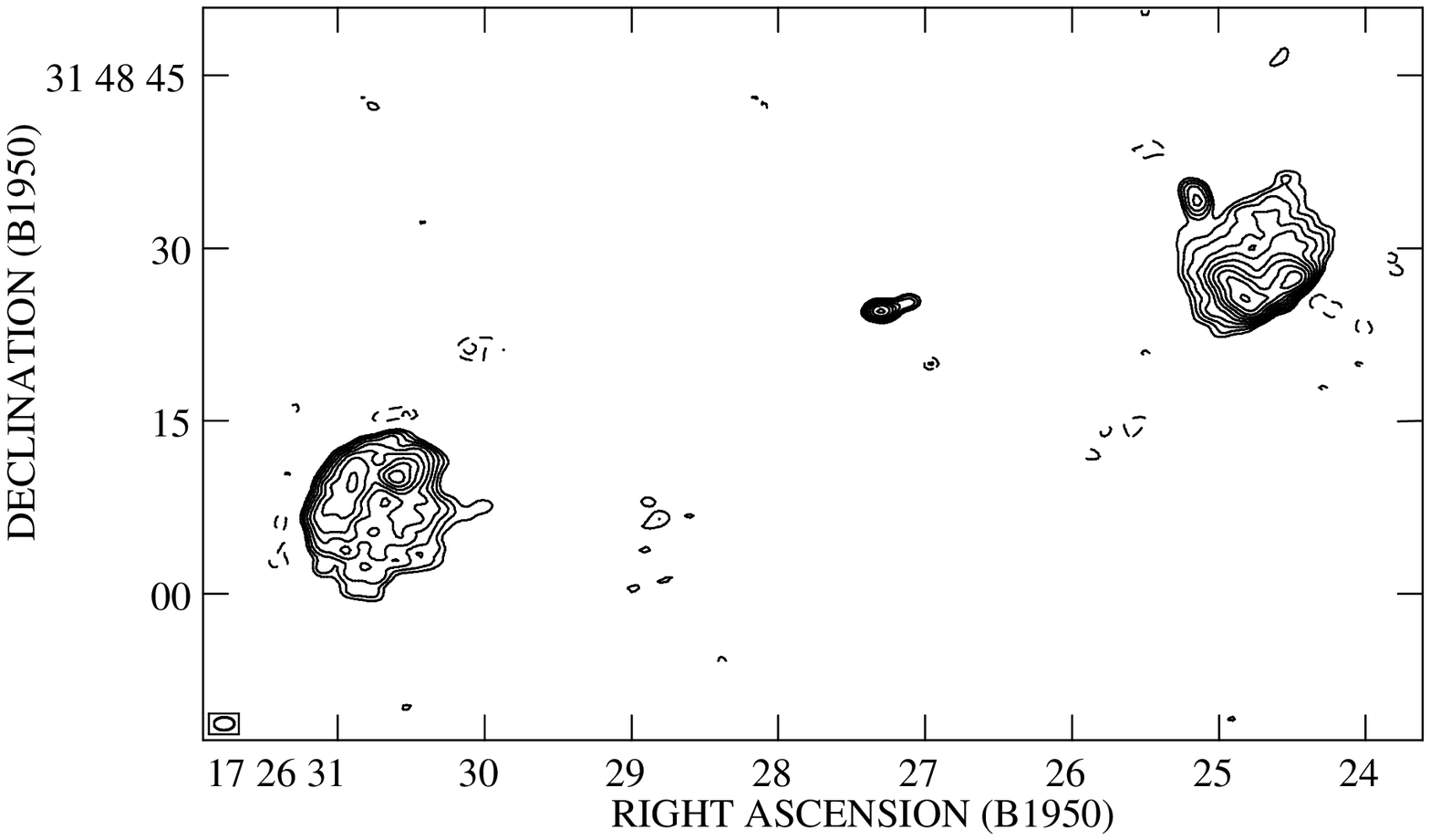}
\figcaption{Total intensity map of 3C\,357 at 4.86 GHz.
The contour levels are $0.5 \times (-\protect\sqrt 2, -1, 1,
\protect\sqrt 2, 2, 2\protect\sqrt 2, \dots)$ mJy beam$^{-1}$. The
peak flux occurs in the western lobe.}
\end{figure}

\begin{figure}
%\epsscale{0.5}
\plotone{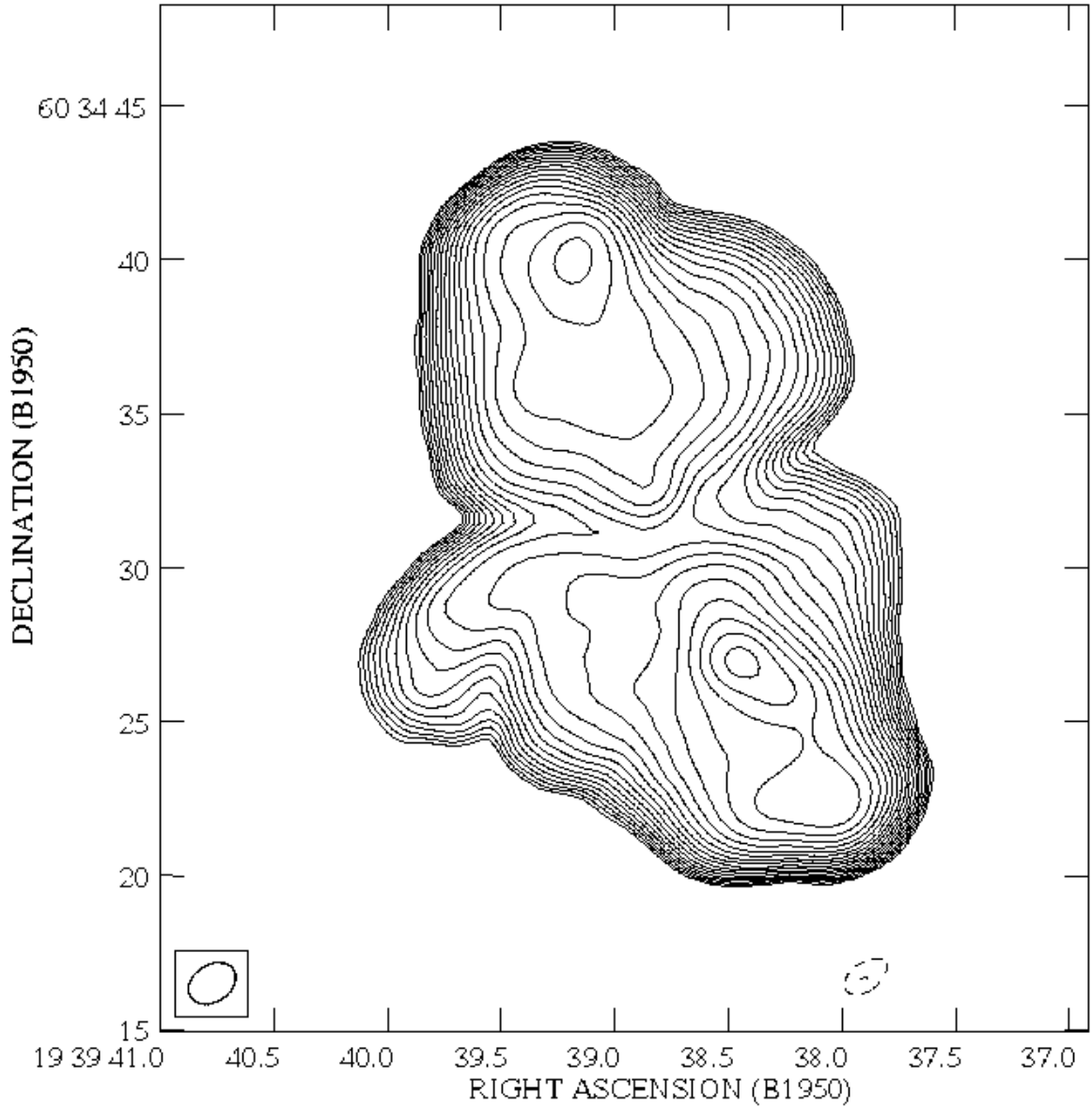}
\figcaption{Total intensity map of 3C\,401 at 1.4 GHz.
The contour levels are $0.4 \times (-\protect\sqrt 2, -1, 1,
\protect\sqrt 2, 2, 2\protect\sqrt 2, \dots)$ mJy beam$^{-1}$. The
peak flux occurs in the jet in the southern lobe.}
%\epsscale{1.0}
\end{figure}
\clearpage
%The above clearpage is there because of LaTeX's limit on unprocessed
%floats. It forces LaTeX to write out the floats it's not yet
%positioned.

\begin{figure}
%\epsscale{0.5}
\plotone{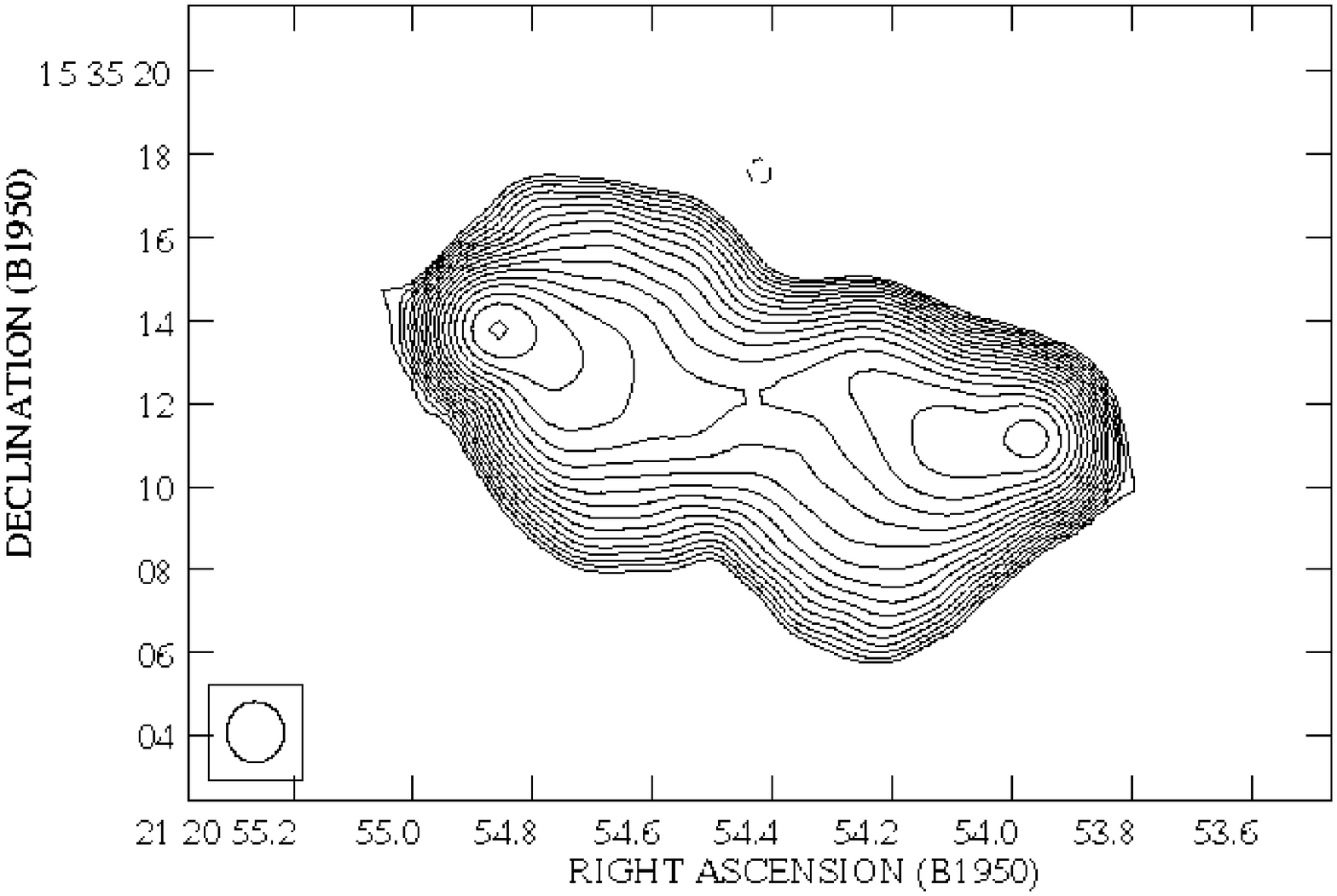}
\figcaption{Total intensity map of 3C\,434 at 1.4 GHz.
The contour levels are $0.5 \times (-\protect\sqrt 2, -1, 1,
\protect\sqrt 2, 2, 2\protect\sqrt 2, \dots)$ mJy beam$^{-1}$. The
peak flux occurs in the eastern lobe.}
%\epsscale{1.0}
\end{figure}

\begin{figure}
\plottwo{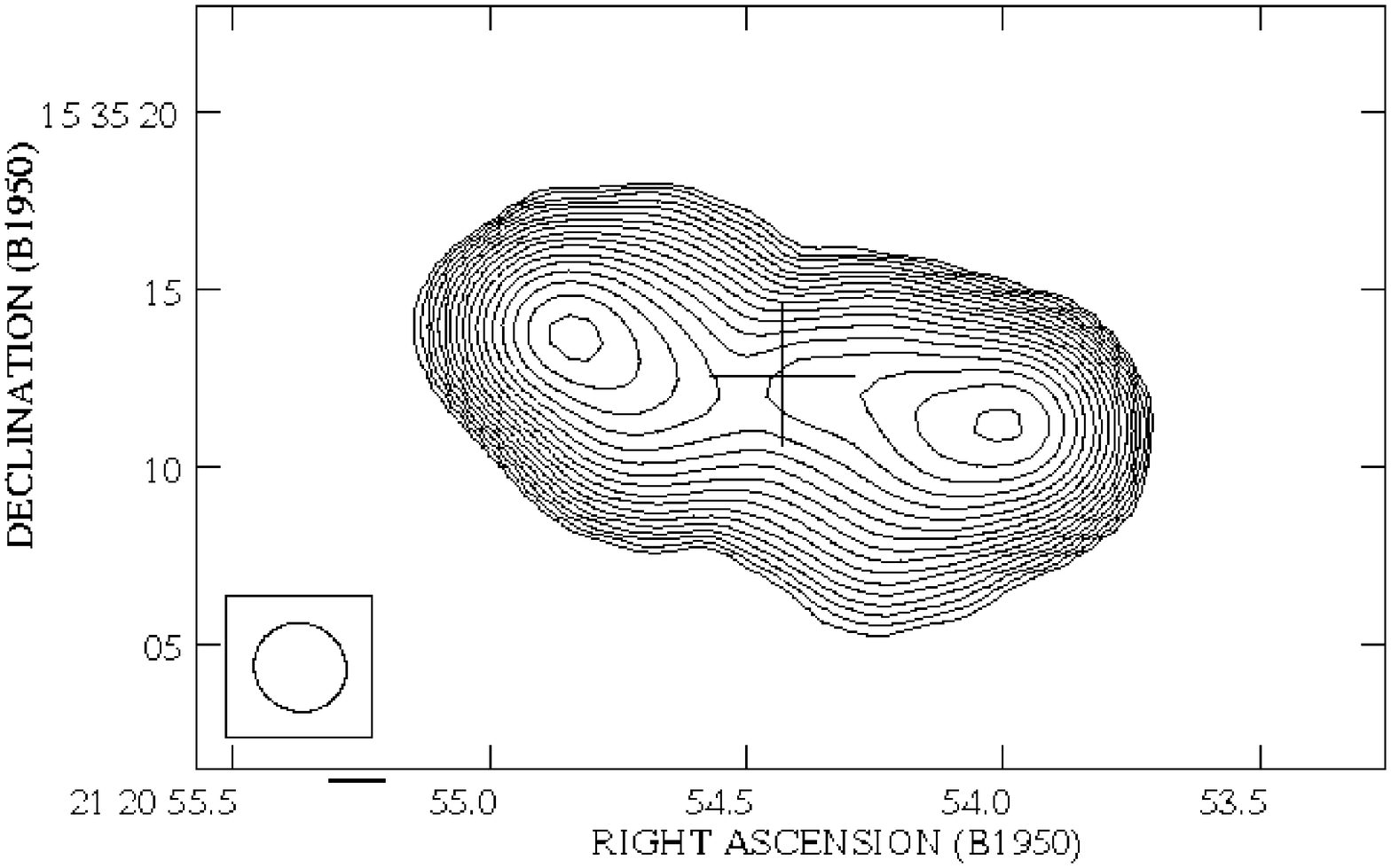}{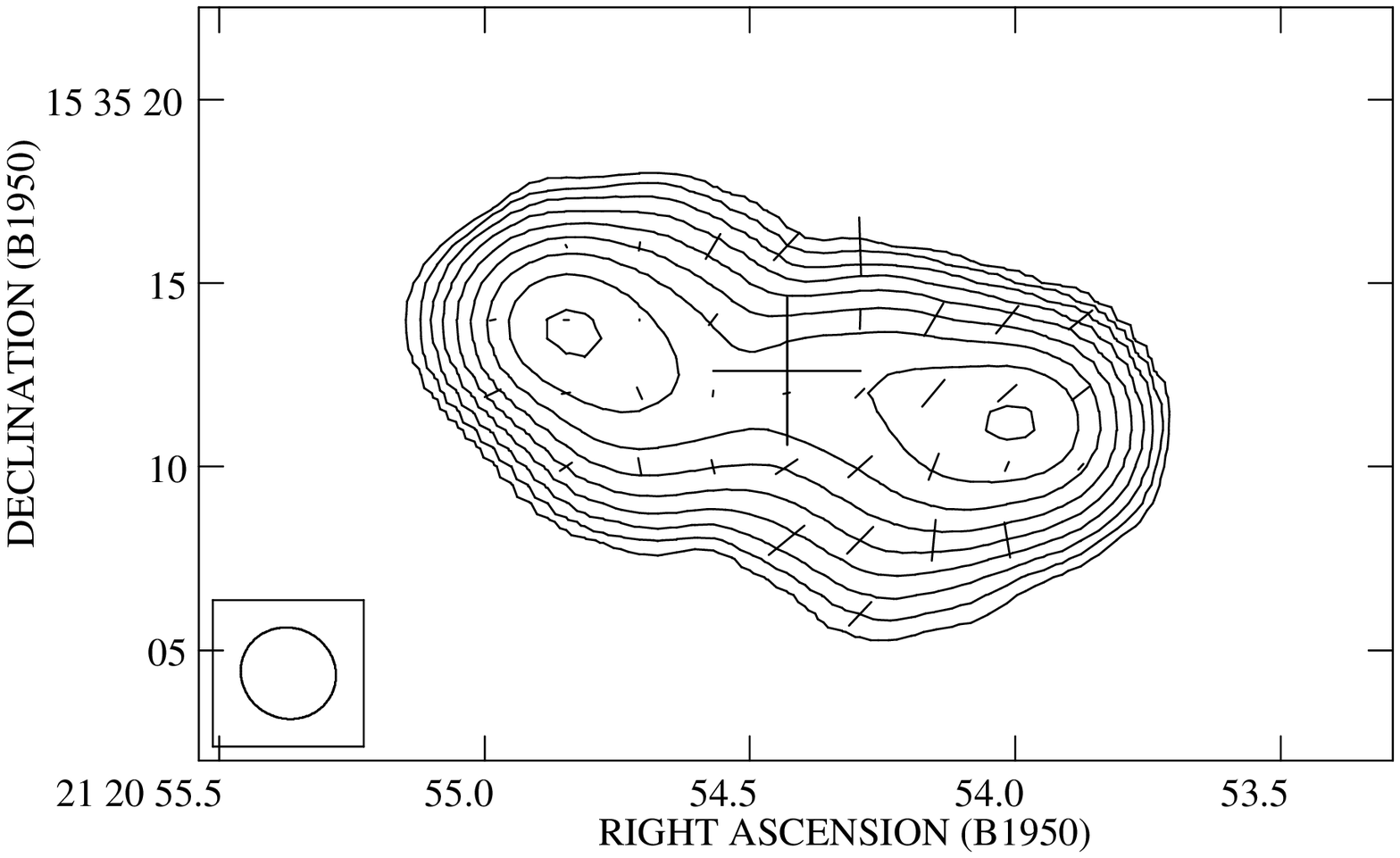}
\figcaption{Left: Total intensity map of 3C\,434 at 8.44 GHz.
The contour levels are $0.2 \times (-\protect\sqrt 2, -1, 1,
\protect\sqrt 2, 2, 2\protect\sqrt 2, \dots)$ mJy beam$^{-1}$. Right:
Polarization map. The contour levels are $0.2 \times (-2, -1, 1, 2, 4,
\dots)$ mJy beam$^{-1}$. A vector of length one arcsecond corresponds
to 20\% polarization. A cross marks the position of the probable core
discussed in the text. The peak flux occurs in the eastern lobe.}
\end{figure}

\begin{figure}
\epsscale{0.75}
\plotone{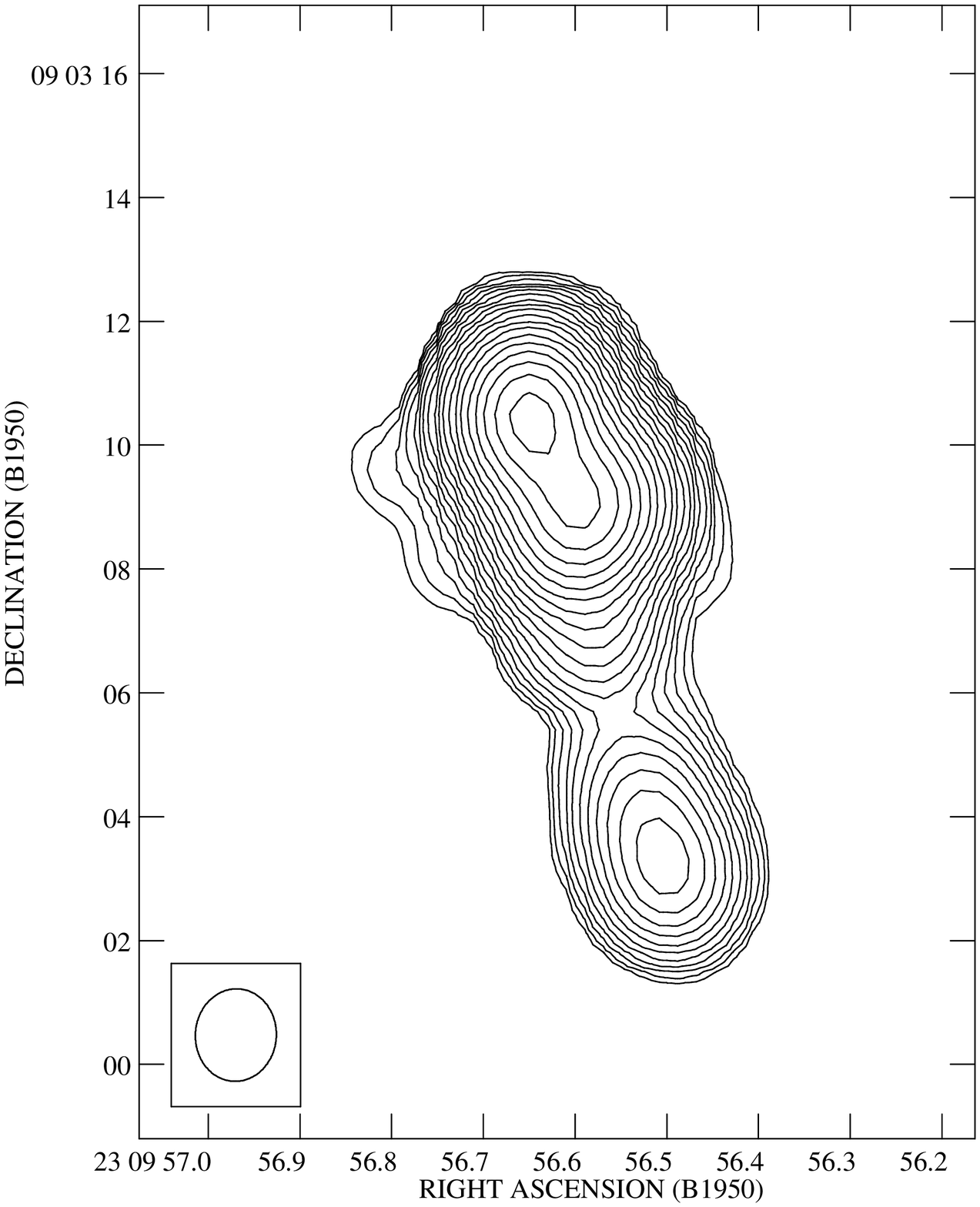}
\figcaption{Total intensity map of 3C\,456 at 1.4 GHz.
The contour levels are $2.0 \times (-\protect\sqrt 2, -1, 1,
\protect\sqrt 2, 2, 2\protect\sqrt 2, \dots)$ mJy beam$^{-1}$. The
peak flux occurs in the northern lobe.}
\epsscale{1.0}
\end{figure}

\begin{figure}
\epsscale{0.50}
\plotone{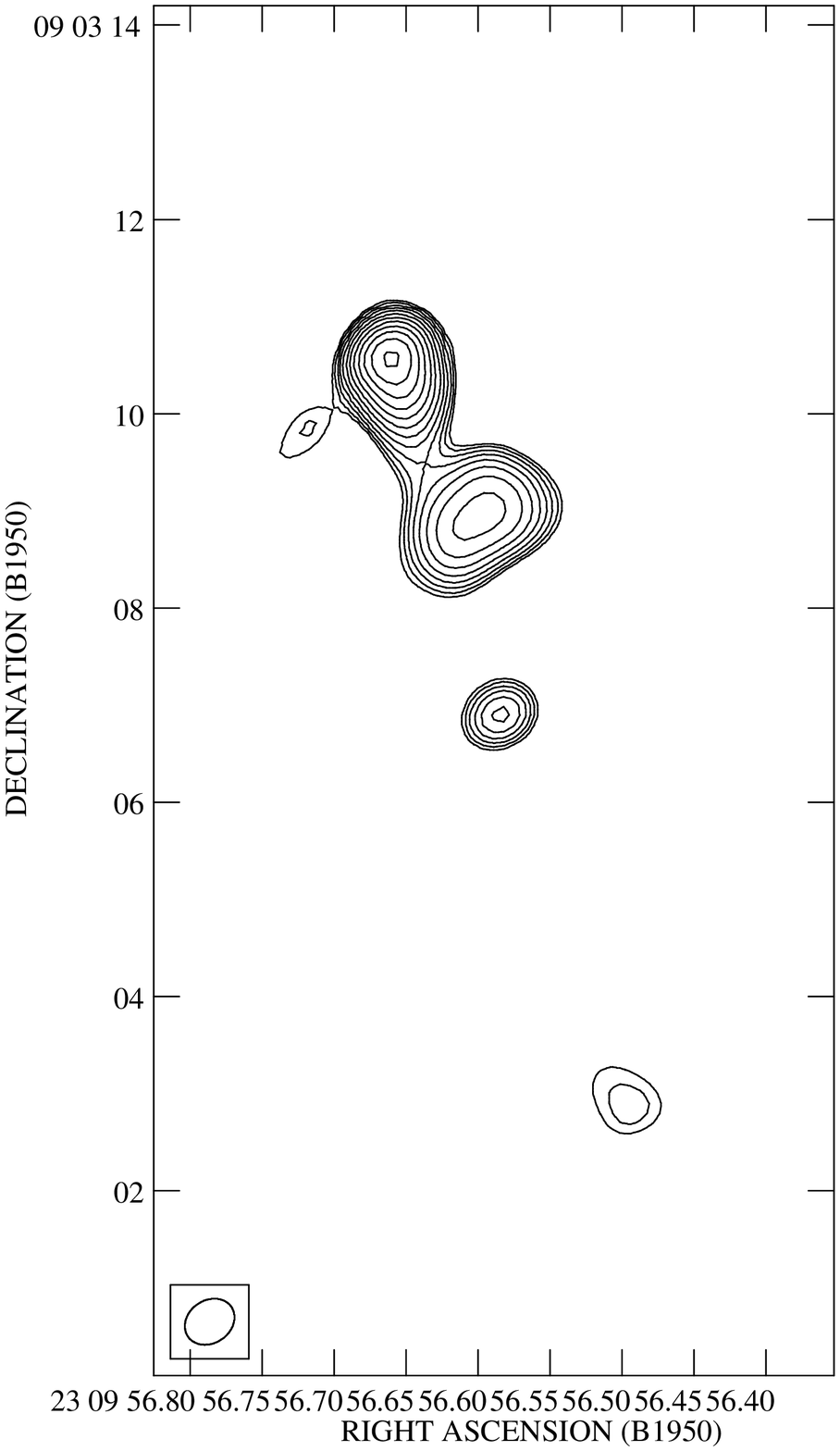}
\figcaption{Total intensity map of 3C\,456 at 4.86 GHz.
The contour levels are $5.0 \times (-\protect\sqrt 2, -1, 1,
\protect\sqrt 2, 2, 2\protect\sqrt 2, \dots)$ mJy beam$^{-1}$. The
peak flux occurs in the northern lobe.}
\epsscale{1.0}
\end{figure}

\begin{figure}
%\epsscale{0.5}
\plotone{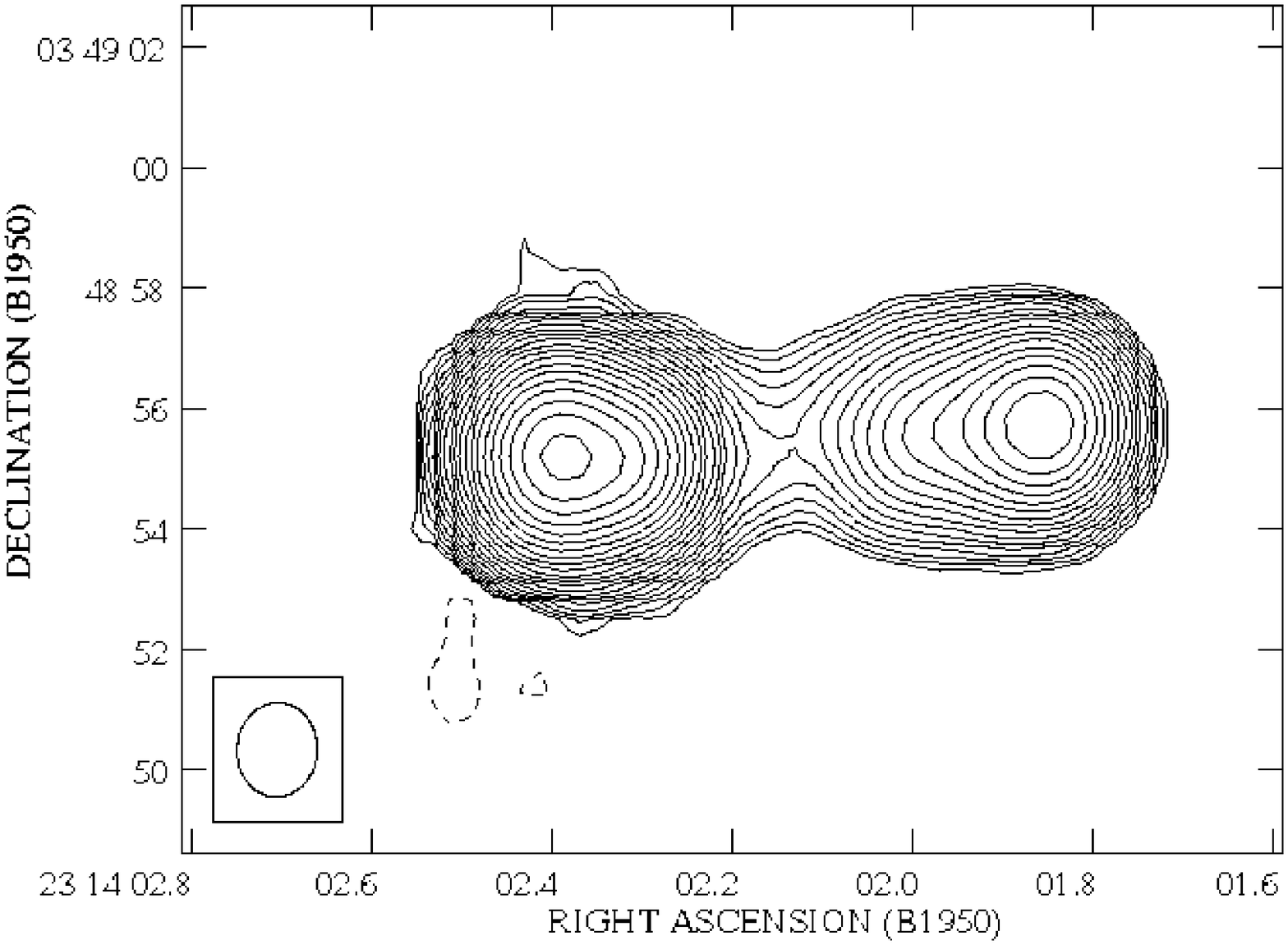}
\figcaption{Total intensity map of 3C\,459 at 1.4 GHz.
The contour levels are $2.0 \times (-\protect\sqrt 2, -1, 1,
\protect\sqrt 2, 2, 2\protect\sqrt 2, \dots)$ mJy beam$^{-1}$. The
peak flux occurs in the eastern lobe.}
%\epsscale{1.0}
\end{figure}

\clearpage

%
%\clearpage
%
%\begin{figure}
%\plotone{sgi9259.eps}
%\caption{We use the \LaTeX\ {\tt figure} environment syntax to set this
%figure caption.}
%\end{figure}

% That's all, folks.

\end{document}